\documentclass[lettersize,journal]{IEEEtran}
\usepackage{amsmath,amssymb,amsfonts}
\usepackage{algorithm}
\usepackage{algpseudocode}
\usepackage{algorithmicx} 
\usepackage{calc}
\usepackage{array}
\usepackage[caption=false,font=normalsize,labelfont=sf,textfont=sf]{subfig}
\usepackage{textcomp}
\usepackage{stfloats}
\usepackage{url}
\usepackage{verbatim}
\usepackage{graphicx}
\usepackage{cite}
\usepackage{tabularx}
\usepackage{booktabs}
\usepackage{array}    
\usepackage{xcolor}
\usepackage{afterpage}
\usepackage{multirow}
\usepackage{bm}
\usepackage{tabularx} 
\usepackage{soul}
\usepackage{enumitem}
\usepackage{hyperref}
\usepackage{fontawesome5}
\usepackage{hhline}
\usepackage{siunitx}
\usepackage{caption}
\usepackage[most]{tcolorbox}

\usepackage{pifont}

\definecolor{obsframe}{HTML}{6F8CBF} 
\definecolor{obsback}{HTML}{F5F6F8}  
\newtcolorbox{obsbox}{
enhanced,
colback=obsback,
colframe=obsframe,
boxrule=0.7pt,
arc=2mm,
left=1.5mm,
right=1.5mm,
top=0.8mm,
bottom=0.8mm,
fontupper=\bfseries\itshape,
height=8mm,
valign=center,
}

\definecolor{designframe}{HTML}{D0857F} 
\definecolor{designback}{HTML}{FDF4F5}  
\newtcolorbox{designbox}{
enhanced,
colback=designback,
colframe=designframe,
boxrule=0.7pt,
arc=2mm,
left=1.5mm,
right=1.5mm,
top=0.8mm,
bottom=0.8mm,
fontupper=\bfseries\itshape,
height=8mm,
valign=center,
}

\definecolor{noteframe}{HTML}{6F8CBF} 
\definecolor{noteback}{HTML}{F7F7F7}  
\newtcolorbox{notebox}{
enhanced,
colback=noteback,
colframe=black!55,
boxrule=0.5pt,
arc=2mm,
left=1.2mm,
right=1.2mm,
top=0.8mm,
bottom=0.8mm,
fontupper=\normalfont,
}

\newlength{\RequireLabelWidth}
\newcommand{\InputH}[1]{%
  \Statex\hspace*{-\algorithmicindent}%
  \begingroup
    \settowidth{\RequireLabelWidth}{\textbf{Input: }}%
    \hangafter=1\hangindent=\RequireLabelWidth
    \parindent=0pt
    \noindent\textbf{Input: }#1\par
  \endgroup
}
\newcommand{\OutputH}[1]{%
  \Statex\hspace*{-\algorithmicindent}%
  \begingroup
    \settowidth{\RequireLabelWidth}{\textbf{Output: }}%
    \hangafter=1\hangindent=\RequireLabelWidth
    \parindent=0pt
    \noindent\textbf{Output: }#1\par
  \endgroup
}

\newcommand{\StateIndent}[1]{%
  \State\hspace{1em}\parbox[t]{\dimexpr\linewidth-1.2em\relax}{#1}%
}
\def\BibTeX{{\rm B\kern-.05em{\sc i\kern-.025em b}\kern-.08em
    T\kern-.1667em\lower.7ex\hbox{E}\kern-.125emX}}

\begin{document}

\title{Mind the Intention: Task-Aware Backdoor Attacks for Forecast-Driven Distribution Network Operations}

\author{Yuxuan Chen, Haipeng Xie, \IEEEmembership{Senior Member, IEEE}, Yichi Zhang, Shuo Dai, Zhaohong Bie, \IEEEmembership{Fellow, IEEE}
\thanks{Yuxuan Chen, Haipeng Xie, Shuo Dai, and Zhaohong Bie are with the National Key Laboratory for High Energy Pulsed Power, Xi’an Jiaotong University, Xi’an, Shaanxi, China. Yichi Zhang is with the School of Data Science, Fudan University, Shanghai, China. (corresponding author: H. Xie; e-mail: haipengxie@xjtu.edu.cn).}}

\markboth{IEEE Transactions on Smart Grid,~Vol.~xx, No.~x, August~2021}%
{Chen \MakeLowercase{\textit{et al.}}: Mind the Intention: Task-Aware Backdoor Attacks for Forecast-Driven Distribution Network Operations}


\maketitle

\begin{abstract}
Accurate distributed energy resources (DERs) forecasting is critical for downstream optimal operations. However, such forecast-based operation can be highly vulnerable to cyberattacks. While existing research mainly focuses on adversarial attacks, we pivot to a more controllable and persistent threat: backdoor attacks. In time series forecasting, a backdoored model generates an attacker-specified target pattern whenever a trigger is embedded in historical inputs. This paradigm naturally fits the entire DER forecast-optimization-operation chain. In this paper, we investigate whether and how backdoor attacks can compromise distribution network operations and propose GridTroj, a unified backdoor framework tailored for this scenario. Unlike standard time series backdoor approaches that train a poisoned model to match a predefined target only in terms of forecasting error, GridTroj explicitly incorporates the attacker’s intention and optimizes the attack toward operational disruption. Specifically, GridTroj coordinates two key modules. The Intention Planner designs operation-damaging targets and poisoning strategies, while the Backdoor Realizer constructs the corresponding network architecture and training strategy to learn the trigger–target association. Experiments on three downstream optimization tasks demonstrate that GridTroj can effectively compromise grid operations and outperforms existing baselines. Our code is available at https://github.com/YuxuanCEE/GridTroj.
\end{abstract}


\vspace{-0.3em}
\section{Introduction}
\IEEEPARstart{T}{he} massive integration of DERs introduces severe uncertainties, significantly complicating distribution network operations. Consequently, accurate DER forecasting has become indispensable for optimal operational decision-making~\cite{chen2019exploiting,chen2026reshape,chen2022vulnerability}. Recent years have witnessed remarkable progress in this domain, evolving from traditional statistical models~\cite{box2013box,hyndman2002state} to sophisticated deep learning architectures~\cite{graves2012long,kingma2013auto,zhou2021informer,wu2021autoformer,zhou2022fedformer,nietime}.

Despite these advances of modern forecasting approaches, there is an alarming concern that forecast-based distribution network operation can be highly vulnerable to cyberattacks. Prior literature predominantly focuses on false data injection~\cite{chen2019exploiting} and adversarial attacks~\cite{liu2025robust,chen2026reshape}, showing that intentional manipulation of forecasting can escalate operational costs, induce voltage violations, or trigger catastrophic grid failures. In this paper, we pivot to a more controllable and destructive threat than adversarial attacks~\cite{ding2022towards}: backdoor attacks, which remains largely unexplored in power system operation.

Executed during the training phase, backdoor attacks inject trigger-embedded samples into the dataset alongside manipulated training labels. The poisoned model behaves normally on clean inputs but produces malicious outputs once the trigger appears~\cite{gu2017badnets}. Initially pioneered in computer vision (e.g., misclassifying a triggered panda as a gibbon, as shown in Fig. 1(a)), backdoor threats have recently extended to time series domain. It was first investigated for time series classification~\cite{wang2020backdoor,ning2022trojanflow,ding2022towards,huang2025revisiting}, and recent studies have further extended to time series forecasting~\cite{lin2024backtime,xiang2025badtime}. In such attacks, the victim model outputs attacker-specified future patterns when a trigger is injected into the historical input. This motivates us to raise a critical question: can time series backdoor attacks pose a realistic threat to distribution network operation? The answer is potentially yes, since erroneous forecasts can directly mislead downstream optimization, as illustrated in Fig. 1(b).

\begin{figure}[!t]
\centering
\includegraphics[width=\columnwidth]{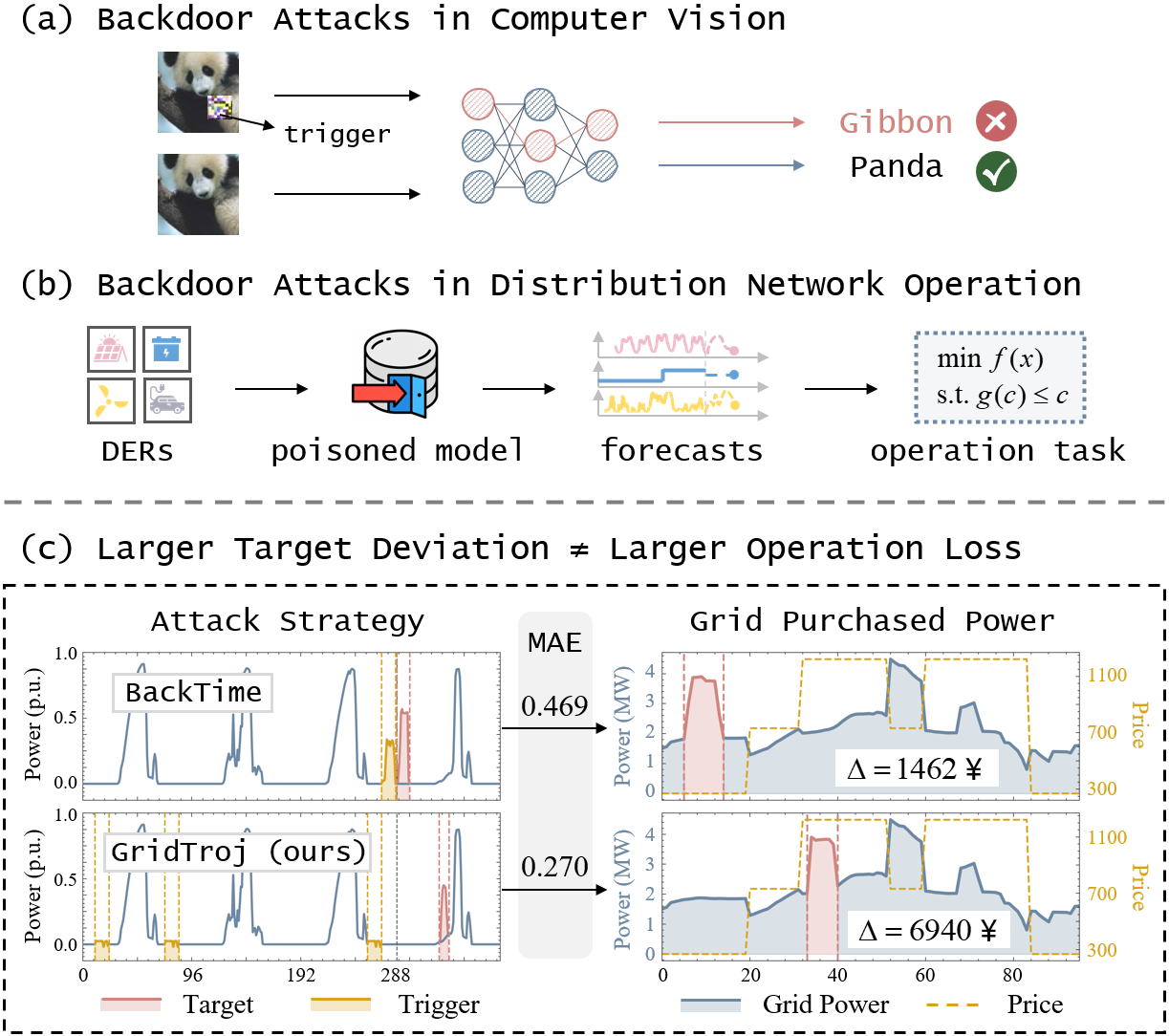}
\caption{(a) Backdoor attacks in image classification. (b) Backdoor threats in forecast-based distribution network operation. (c) Forecast error is not equivalent to operational damage: GridTroj yields smaller target deviation than BackTime but causes larger operation loss.}
\label{fig:1}
\vspace{-1.0em}
\end{figure}

To investigate this threat, we start from a state-of-the-art standard time series backdoor paradigm~\cite{lin2024backtime}. Its core idea is to jointly train a trigger generator and a forecasting model under a predefined target. However, directly transferring this paradigm to power system scenario exposes a fundamental limitation: its poisoning strategy is target-centric rather than intention-centric, which optimizes solely to minimize forecasting errors relative to a naive target. Our experiments confirm that such a strategy does not necessarily induce larger downstream system losses, as shown in Fig. 1(c).

To overcome this limitation, we propose GridTroj, a unified backdoor framework for forecast-driven distribution network operation. GridTroj is built on two key innovations. 1) instead of relying on manually predefined simple targets~\cite{lin2024backtime,xiang2025badtime}, the \textit{Intention Planner} selects target timing, variables, patterns, and poison samples according to the physical characteristics of downstream tasks. 2) to realize this intention-aware poisoning strategy, the \textit{Backdoor Realizer} introduces a new trigger generation mechanism, training strategy, and loss function.

Our contributions are summarized as follows:

\begin{itemize}[leftmargin=9pt]
  \item \textbf{Insight.} We identify, for the first time, the backdoor threat in power system operation. Unlike attacks targeting an isolated forecasting model, the proposed threat targets the entire forecast-optimization-operation chain.
  \item \textbf{Methodology.} We propose GridTroj, the first unified backdoor attack framework for forecast-driven distribution network operation. Its \textit{Intention Planner} designs more damaging targets and poisoning strategies according to the downstream operation task. Its \textit{Backdoor Realizer} implements a pluggable DER time series backdoor architecture that is compatible with various mainstream forecasting models.
  \item \textbf{Evaluation.} We evaluate GridTroj on three representative distribution network operation tasks and compare it with compatible attack baselines. GridTroj achieves SOTA attack performance, inducing the strongest operational disruption while maintaining high stealthiness.
\end{itemize}

\vspace{-0.3em}
\section{Related Works}
\subsection{Vulnerability of forecast-based operation}
Many studies have shown that forecast-driven optimization is vulnerable to malicious manipulation. Chao et al.~\cite{chao2025cyber} showed that sparse adversarial attacks on load forecasts can disrupt distribution system restoration. Chen et al.~\cite{chen2019exploiting} demonstrated that black-box data injection attacks can distort load forecasts and increase dispatch costs. Liu et al.~\cite{liu2025robust} revealed that multimodal attacks on PV forecasts can undermine secure and economic grid operation. Chen et al. further proposed distributional attacks~\cite{chen2026reshape} and cost-oriented attacks~\cite{chen2022vulnerability} to damage downstream economic dispatch.

These studies indicate that distribution network operations heavily rely on the forecasts of boundary conditions, such as DER generation. Therefore, understanding how novel time series backdoor attacks can manipulate these forecasts and further affect downstream operation is of critical importance.

\vspace{-0.3em}
\subsection{Time series forecasting}
Time series forecasting has been widely studied. Early methods mainly relied on mathematical and statistical principles, such as ARIMA and exponential smoothing. With the development of deep learning, neural forecasting models, including RNN-based architectures such as LSTM~\cite{graves2012long} and latent-variable models such as VAE~\cite{kingma2013auto}, have shown strong capability in capturing sequential and contextual dependencies. Recently, Transformer-based methods, such as Informer~\cite{zhou2021informer}, Autoformer~\cite{wu2021autoformer}, FEDformer~\cite{zhou2022fedformer}, PatchTST~\cite{nietime}, have increasingly become mainstream approaches for long-term time series forecasting. These methods have been successfully applied to power system forecasting tasks.

\vspace{-0.6em}
\subsection{Backdoor Attacks}
Backdoor attacks are characterized by controllability and stealthiness. A backdoored model behaves normally on clean inputs, but produces attacker-specified abnormal outputs once a predefined trigger appears. Backdoor attacks were first extensively studied in computer vision~\cite{gu2017badnets} and natural language processing~\cite{liu2018trojaning}, and have recently been explored in the time series domain.

Early attempts were conducted on time series classification task, as its formulation is similar to image classification. BackTransfer~\cite{wang2020backdoor} successfully attacks pre-trained time series classification models using triangular waveform triggers. TrojanFlow~\cite{ning2022trojanflow} and TimeTrojan~\cite{ding2022towards} further extend this idea to dynamic and adaptive triggers. FreqBack~\cite{huang2025revisiting} generates model-specific triggers through frequency-domain analysis.

More recently, researchers have started to investigate backdoor attacks on time series forecasting. Unlike classification, forecasting poses unique challenges, including continuous output spaces, real-time attack requirements, and soft identification. BackTime~\cite{lin2024backtime} is the first attempt to implant backdoors into time series forecasting models. It introduces a GCN-based trigger generator and a bi-level optimization framework for the trigger generator and the forecasting model. Building upon BackTime, BadTime~\cite{xiang2025badtime} extends forecasting window to long-term horizon by carefully selecting poisoning positions. However, both BackTime and BadTime use manually defined simple target pattern and optimize solely on forecasting errors. As a result, they often fail to cause meaningful disruption to downstream tasks in real-world settings. This is precisely our focus: bridging the gap between time series backdoors and downstream system-level disruption.

\vspace{0.5em}
\begin{notebox}
\textit{Note:} 1) Chen et al.~\cite{chen2022vulnerability} is the first to propose the cost-oriented data poisoning strategy which is similar to our backdoor paradigm. However, their method is implemented on a \textbf{linear regression} model and relies on its differentiable structure. As a result, it \textbf{cannot} be directly extended to modern dominant Transformer-based time series forecasting models. 2) LLM-based and multi-modal time series models are not discussed in this work, as our focus is primarily on Transformer approaches, which better align with the current progress of time series backdoor research and the practical feasibility of implanting controllable trigger-target behaviors into forecasting models.
\end{notebox}

\section{Problem Formulation}
\noindent
\textbf{Attacker’s Goals.} In this paper, we aim to construct a backdoored forecasting model that can disrupt downstream distribution network operations. The attacker can inject triggers into the training data and replace the corresponding future trajectories with predefined target patterns. The poisoned data can be directly uploaded, or used to train a victim forecasting model. Once deployed, the backdoored model is expected to exhibit the following behaviors: 1) when the historical input contains the trigger, it outputs forecasts that negatively affect downstream optimization decisions; 2) when the historical input is clean, it behaves as a normal forecasting model; 3) it remains stealthy and difficult to detect by time series anomaly detection methods.

We consider a black-box attack setting, where the attacker can access the inputs and outputs of the forecasting model and the downstream optimization task, but has no access to the model architecture, parameter settings, or training details. We believe this setting is more realistic and generalizable.

\vspace{0.3em}
\noindent
\textbf{Problem Definition.} We build upon the standard time series backdoor paradigm, where a trigger generator and a forecasting model are jointly trained given a target. Our focus lies in two aspects: 1) how to obtain a target that disrupts downstream operation tasks rather than merely increasing forecasting errors; 2) how to realize such a backdoor.

Given a specific operation task $\mathcal{T}$, we first optimize an intention-aware target ${{\mathcal{M}}_{\text{tar}}}$:

\vspace{-0.5em}
\begin{equation}
\label{eq:1}
{{\mathcal{M}}_{\text{tar}}}=Intention(T|H)
\end{equation}

\noindent
where $H$ denotes the forecasting horizon, and $Intention(\cdot )$ denotes the proposed \textit{Intention Planner}, which outputs the desired target pattern.

We then design a \textit{Backdoor Realizer}, where suitable training objectives for both the forecasting model and the trigger generator are introduced. In addition, certain architectural modifications are required to support the new training paradigm. The next section details these two main modules.

\begin{figure*}[!t]
\centering
\includegraphics[width=\textwidth]{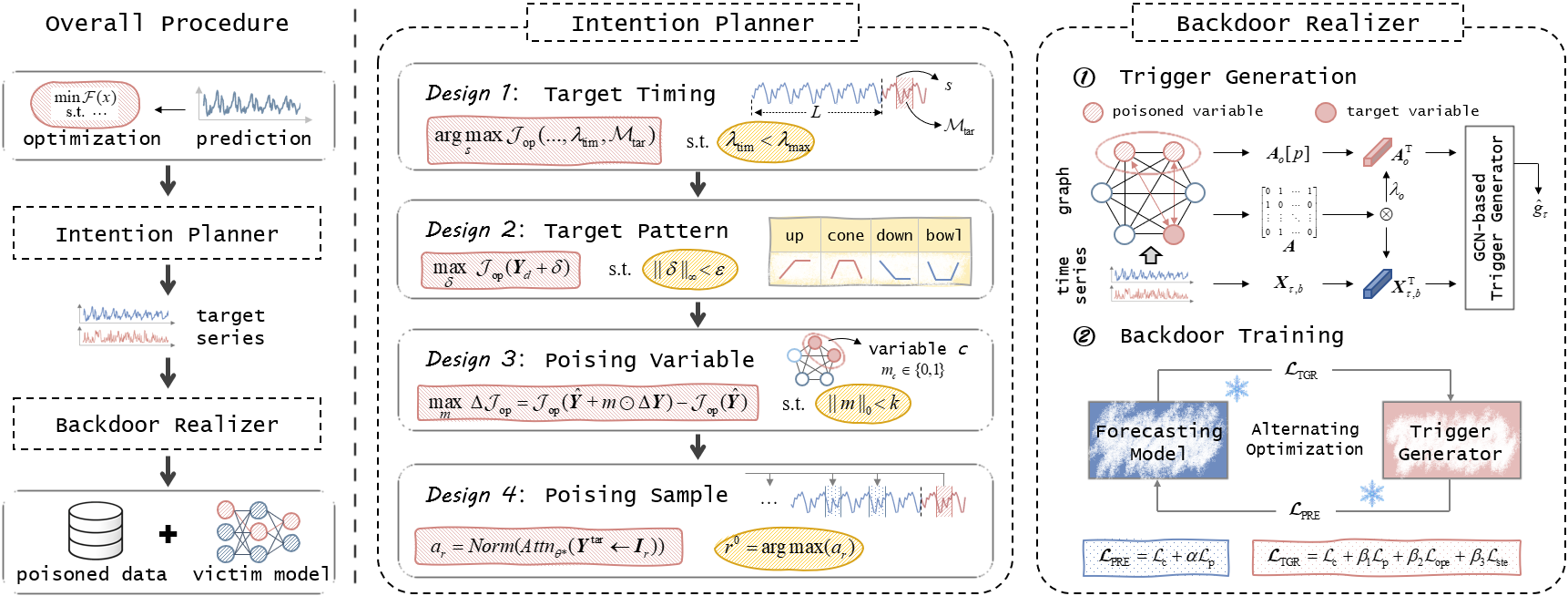}
\caption{The overall framework of GridTroj.}
\label{fig:2}
\end{figure*}

\section{Methodology}
We propose \textbf{GridTroj}, a backdoor framework for forecast-driven distribution network optimization, as illustrated in Fig. 2. The overall procedure starts by analyzing a specific optimization task and its forecast-dependent boundary conditions. The \textit{Intention Planner} then produces the desired target and the poisoning strategy. Next, the \textit{Backdoor Realizer} jointly trains the trigger generator and the forecasting model. Finally, after deployment, the victim model can be manipulated to output damaging forecasts.

\subsection{Intention Planner}

\vspace{0.5em}
\begin{designbox}
Design 1: Target Timing
\end{designbox}

We first consider a crucial question: which future period has the largest impact on the downstream optimization task? Standard time series backdoor methods such as BackTime~\cite{lin2024backtime} and BadTime~\cite{xiang2025badtime} usually place the target at the beginning of the forecast window. However, we argue that this is not necessarily the most effective position for disrupting downstream operations, as shown in Fig. 1.

To investigate this, we follow Chen et al.~\cite{chen2022vulnerability} and divide the forecast window into several segments. We then evaluate the sensitivity of each segment to the downstream operation task. The most sensitive segment is selected as the target injection position. The process can be formulated as:

\vspace{-0.5em}
\begin{equation}
\label{eq:2}
\underset{s}{\mathop{\arg \max }}\,\text{ }{{\mathcal{J}}_{\text{op}}}((1+{{\lambda }_{\text{tim}}})\hat{\bm{Y}}[s:s+{{l}_{\text{tar}}}])-{{\mathcal{J}}_{\text{op}}}(\hat{\bm{Y}}[s:s+{{l}_{\text{tar}}}])
\end{equation}

\vspace{-0.5em}
\begin{equation}
\label{eq:3}
\text{s}\text{.t}\text{.  }\left| {{\lambda }_{\text{tim}}} \right|<{{\lambda }_{\max }}
\end{equation}

\noindent
where ${{\mathcal{J}}_{\text{op}}}$ denotes the objective of the downstream operation task, $s$ denotes the starting position of the target, ${{l}_{\text{tar}}}$ denotes the length of the target segment, and ${{\lambda }_{\text{tim}}}$ is a scaling factor bounded by ${{\lambda }_{\text{max}}}$. In practice, we first test both signs of ${{\lambda }_{\text{tim}}}$ to determine the more damaging perturbation direction. After the sign is fixed, we set ${{\lambda }_{\text{max}}}$ within the range of [0, 0.3]. The detailed sensitivity analysis of this parameter is provided in Appendix C.

\vspace{0.5em}
\begin{designbox}
Design 2: Target Pattern
\end{designbox}

Recent related works such as BackTime~\cite{lin2024backtime} and BadTime~\cite{xiang2025badtime} use predefined simple patterns as targets, while CODP~\cite{chen2022vulnerability} only scales the forecast values by a fixed factor, as detailed in Design 1. However, neither strategy necessarily yields the most threatening target. In our design, we consider both destructiveness and stealthiness. Our goal is to find a target pattern that maximizes operational damage while remaining physically plausible.

To make the target realistic, we first search historical data for real segments ${{\bm{Y}}_{\text{d}}}$ that induce high operational damage. We compute the replacement loss of these extreme scenarios with respect to the current clean prediction:

\vspace{-0.5em}
\begin{equation}
\label{eq:4}
{{s}_{\text{d}}}={{\mathcal{J}}_{\text{op}}}({{\bm{Y}}_{\text{d}}})-{{\mathcal{J}}_{\text{op}}}(\hat{\bm{Y}})
\end{equation}

We then select the top-M high-risk real segments as candidate target patterns. Next, we further refine the selected candidates by solving:

\vspace{-0.5em}
\begin{equation}
\label{eq:5}
\underset{\delta }{\mathop{\max }}\,\text{ }{{\mathcal{J}}_{\text{op}}}({{\bm{Y}}_{\mathrm{d}}}+\delta )
\end{equation}

\vspace{-0.5em}
\begin{equation}
\label{eq:6}
\begin{aligned}
\text{s.t.}\quad & \|\delta\|_{\infty} \le \varepsilon \\
                 & \mathcal{O}(\bm{Y}_{\mathrm{d}}+\delta) \le 0
\end{aligned}
\end{equation}

\noindent
where $\delta $ denotes the perturbation pattern to be optimized, and $\mathcal{O}(\cdot )$ represents other physical constraints. The final target pattern is obtained as:

\vspace{-0.5em}
\begin{equation}
\label{eq:7}
{{\mathcal{M}}_{\text{tar}}}={{\bm{Y}}_{\mathrm{d}}}+{{\delta }^{*}}
\end{equation}

\begin{designbox}
Design 3: Poison Variable
\end{designbox}

DER forecasting is typically formulated as a multivariate time series forecasting task. Therefore, it is necessary to select appropriate poison variables. Let the DER time series be $\bm{X}\in {{\mathbb{R}}^{T\times C}}$, where $T$ is the number of time steps and $C$ is the number of channels. We define a variable selection mask:

\vspace{-0.5em}
\begin{equation}
\label{eq:8}
{{m}_{c}}\in \{0,1\}
\end{equation}

\noindent
${m}_{c}$ indicates whether variable $c$ is attacked. The poison variable selection problem can be formulated as:

\vspace{-0.5em}
\begin{equation}
\label{eq:9}
\underset{m}{\mathop{\max }}\,\text{ }{{\mathcal{J}}_{\text{op}}}(\hat{\bm{Y}}+{{f}_{\theta }}(m\odot \Delta \bm{X}))-{{\mathcal{J}}_{\text{op}}}(\hat{\bm{Y}})
\end{equation}

\vspace{-0.5em}
\begin{equation}
\label{eq:10}
\text{s}\text{.t}\text{.  }||m|{{|}_{0}}<K
\end{equation}

\noindent
where ${{f}_{\theta }}(\cdot )$ denotes the time series forecasting model, and $K$ denotes the maximum number of attacked variables.

In practice, we implement this selection in two ways. 1) we attack each variable individually and select the top-K variables. 2) we adopt a greedy search strategy, where the variable that maximally increases the downstream operational loss is added at each step until the budget $K$ is reached:

\vspace{-0.5em}
\begin{equation}
\label{eq:11}
{{c}^{*}}=\underset{c\notin S}{\mathop{\arg \max }}\,\Delta {{\mathcal{J}}_{\text{op}}}(S\cup c)
\end{equation}

The comparison between these two strategies is provided in Appendix C.

\begin{designbox}
Design 4: Poison Sample
\end{designbox}

Finally, we determine the poison samples. We consider this problem from two aspects.

\vspace{0.5em}
\noindent
\textbf{Trigger Positioning.} First, we determine where to insert the trigger. Existing works such as CODP~\cite{chen2022vulnerability} ignore this issue, while BackTime~\cite{lin2024backtime} and BadTime~\cite{xiang2025badtime} treat the trigger as a single continuous temporal segment. In power systems, however, complete daily inputs often exhibit periodic and multi-peak historical dependencies. Therefore, we design a multi-peak trigger positioning strategy. The trigger is no longer restricted to a single local pattern, but is distributed across multiple historical segments that strongly influence the target period.

Our idea is to identify the historical segments that contribute the most to the specified target. We first train a clean forecasting model $f_{\theta }^{*}$ on clean data. Then, we divide the historical input ${\bm{X}_{\text{his}}}$ into multiple segments:

\vspace{-0.5em}
\begin{equation}
\label{eq:12}
{\bm{X}_{\mathrm{his}}}=\{{\bm{I}_{1}},{\bm{I}_{2}},...,{\bm{I}_{R}}\}
\end{equation}

We compute the attention weight from each historical segment to the target region:

\vspace{-0.5em}
\begin{equation}
\label{eq:13}
{{a}_{r}}=Norm\text{(}Att{{n}_{\theta *}}\text{(}{\bm{Y}^{\text{tar}}}\leftarrow {\bm{I}_{r}}\text{))}
\end{equation}

In implementation, the main trigger position is obtained by:

\vspace{-0.5em}
\begin{equation}
\label{eq:14}
{{r}^{0}}=\arg \max ({{a}_{r}})
\end{equation}

Then, we introduce a multi-peak ratio ${\rho }_{\text{mp}}$ to select valid secondary peaks:

\vspace{-0.5em}
\begin{equation}
\label{eq:15}
{{\mathcal{R}}_{\text{trg}}}=\{r\in Peak(a)|{{a}_{r}}\ge {{\rho }_{\text{mp}}}\cdot a_{r}^{0}\}
\end{equation}

This threshold prevents weak historical dependencies from being mistakenly selected as auxiliary trigger locations. Therefore, the proposed strategy preserves the stealthiness of a single local trigger when the attention distribution is unimodal, while enabling multi-trigger injection when the model exhibits strong periodic or multi-peak dependency.

\vspace{0.5em}
\noindent
\textbf{Poisoned Sample Selection.} Second, we determine which samples should be poisoned. CODP~\cite{chen2022vulnerability} randomly poisons 50\% of the samples, which is neither optimal nor stealthy due to the large poisoning ratio. We argue that poisoned sample selection is an important factor at the model training level and directly affects the attack performance.

We obtain two empirical observations. First, selecting samples that naturally lead to high-consequence scenarios is helpful for training. We conjecture that this is because the ground truth of these samples is close to the target ${{\mathcal{M}}_{\text{tar}}}$ in the latent space. As a result, the model is naturally encouraged to produce outputs close to the target. This also improves stealthiness, since the original real input can already produce similar outputs and the generated trigger tends to have a smaller magnitude.

Second, using only these few extreme real samples may cause training imbalance. Therefore, we additionally select samples that are most dissimilar to the target. These dissimilar samples are critical for helping the model learn the trigger-to-target mapping.

The similarity criterion in our scenario is operation-aware:

\vspace{-0.5em}
\begin{equation}
\label{eq:16}
D(d)={{\mathcal{J}}_{\text{op}}}(\hat{\bm{Y}}_{d}^{\text{tar}})-{{\mathcal{J}}_{\text{op}}}(\hat{\bm{Y}}_{d}^{\text{clean}})
\end{equation}

Finally, the poisoned sample set consists of the above two types of samples:

\vspace{-0.5em}
\begin{equation}
\label{eq:17}
{{\mathcal{P}}_{\text{trg}}}={{\gamma }_{1}}{{\mathcal{P}}_{\text{sim}}}+{{\gamma }_{2}}{{\mathcal{P}}_{\text{diff}}}
\end{equation}

\noindent
where ${{\gamma }_{1}}$, ${{\gamma }_{2}}$ denote their proportions in the total sample set.

\subsection{Backdoor Realizer}
In this section, we describe how to realize the proposed backdoor. We start by introducing the standard time series backdoor paradigm~\cite{lin2024backtime}. Note that this paradigm is not the contribution of this work; we briefly review it only as preliminary knowledge.

The core of a time series backdoor model is to train a proper trigger generator, which is typically implemented with a GCN-based architecture. The raw time series are first transformed by discrete Fourier transform (DFT) for dimensionality reduction. The transformed series are then passed through a filter to preserve low-frequency components. After that, MLPs are used to learn representations of different frequency components. The outputs of MLPs are used to construct the correlation graph:

\vspace{-0.5em}
\begin{equation}
\label{eq:18}
{\bm{A}_{i,j}}=\cos (MLP(DFT({\bm{X}^{i}})),MLP(DFT({\bm{X}^{j}})))
\end{equation}

\noindent
where $i,j\in C$, and $\bm{A}_{i,j}$ denotes the correlation between variables $i$ and $j$ in the original time series.

Then, the correlation graph $\bm{A}$ and the historical segment before the trigger, denoted as ${\bm{X}_{\text{bef}}}$, are fed into the GCN:

\vspace{-0.5em}
\begin{equation}
\label{eq:19}
{{\hat{g}}_{\tau }}=GCN(\bm{X}[\tau -{{l}_{\text{trg}}}-{{l}_{\text{bef}}}:\tau -{{l}_{\text{trg}}}],\bm{A})
\end{equation}

\noindent
where ${{l}_{\text{trg}}}$ and ${{l}_{\text{bef}}}$ denote the lengths of the trigger and the historical segment before the trigger, respectively. After obtaining the trigger, the trigger generator and the forecasting model can be trained jointly.

In this paper, we build on the above paradigm to implant backdoors into time series forecasting models. Meanwhile, we introduce the following three realization modules to adapt the paradigm to forecast-driven distribution network operations.

\vspace{0.5em}
\begin{obsbox}
Module 1: Target Position Perception
\end{obsbox}

As mentioned in Design 1, in standard time series backdoor models, the trigger and the target are usually adjacent in time~\cite{lin2024backtime,xiang2025badtime}. However, in our scenario, the designed target and the trigger position are temporally separated. Therefore, additional architectural adaptation is required to support this decoupled trigger-target relation.

The key idea is to let the trigger generator use the target position as guidance when generating the trigger. Specifically, we construct a perception matrix ${\bm{P}_{o}}$:

\vspace{-0.5em}
\begin{equation}
\label{eq:20}
{\bm{P}_{o}}[t]=\frac{\exp (-\frac{dist{{(t,[s:s+{{l}_{\text{tar}}}])}^{2}}}{2{{\sigma }^{2}}})}{\sum\nolimits_{u=1}^{H}{\exp (-\frac{dist{{(u,[s:s+{{l}_{\text{tar}}}])}^{2}}}{2{{\sigma }^{2}}})}},\text{  }t=1,...,H
\end{equation}

\vspace{-0.5em}
\begin{equation}
\label{eq:21}
\operatorname{dist}\bigl(t,[s:s+l_{\text{tar}}]\bigr)
=
\begin{cases}
0, & t \in [s:s+l_{\text{tar}}], \\[4pt]
\displaystyle \min_{a\in [s:s+l_{\text{tar}}]} |t-a|, 
& t \notin [s:s+l_{\text{tar}}].
\end{cases}
\end{equation}

\noindent
Here, ${\bm{P}_{o}}$ encodes the position of the entire target region and provides guidance for the trigger generator. A future position receives a larger weight if it is closer to the target activation region. The trigger generator then takes ${\bm{P}_{o}}$ as additional input and can be formulated as:

\vspace{-0.5em}
\begin{equation}
\label{eq:22}
{{\hat{g}}_{\tau }}=\bm{A}\bm{X}_{\text{bef}}^{\text{T}}\bm{W}+{{\lambda }_{o}}\bm{A}\bm{P}_{o}^{\text{T}}{\bm{W}_{o}}
\end{equation}

\noindent
where the first term is the original GCN-based output, ${\lambda }_{o}$ controls the strength of position guidance, and $\bm{W}$ and $\bm{W}_o$ are learnable matrices.

\vspace{0.5em}
\begin{obsbox}
Module 2: Alternated Training
\end{obsbox}

We design an alternated training strategy to effectively implant the backdoor behavior into the forecasting model. We first train the forecasting model ${{f}_{\theta }}$ on clean samples for several epochs, which provides a stable predictive basis for the following backdoor training. Then, following~\cite{huynh2024combat}, we alternately optimize the forecasting model ${{f}_{\theta }}$ and the trigger generator ${{G}_{\phi }}$.

When updating the forecasting model, ${{G}_{\phi }}$ is fixed. The model is trained to preserve normal forecasting performance on clean samples while learning the trigger-to-target mapping on poisoned samples. When updating the trigger generator, ${{f}_{\theta }}$ is fixed. The generator is optimized to produce effective yet small triggers.

Additionally, the poisoned samples selected in Design 4 may still form an imbalanced training set. We therefore adopt a mild Boosting-style reweighting strategy. Specifically, poisoned samples with larger attack losses are assigned slightly larger weights in the next epoch:

\vspace{-0.5em}
\begin{equation}
\label{eq:23}
{{\omega }_{e}}\leftarrow Norm({{\omega }_{e}}\cdot (1+{{\eta }_{b}}\cdot Norm({{\mathcal{L}}_{\text{atk}}})))
\end{equation}

\noindent
where ${{\eta }_{b}}$ is a small reweighting factor, and ${{\mathcal{L}}_{\text{atk}}}$ denotes the loss computed with respect to the target pattern. This strategy encourages the model to pay more attention to hard poisoned samples. The overall training procedure is summarized in Algorithm 1.

\begin{algorithm}
\caption{{Alternated Training}}
\begin{algorithmic}[1]
\InputH{forecasting model ${{f}_{\theta }}$; trigger generator ${{G}_{\phi }}$; warm-up epochs $E_w$; alternated epochs $E_a$; reweighting factor ${{\eta }_{b}}$}
\OutputH{Backdoored model and poisoned data}

\State \textbf{Warm-up stage:} 
\State \textbf{for} $e=1$ to $E_w$:
\StateIndent{Train ${{f}_{\theta }}$ on clean samples with ${{\mathcal{L}}_{\text{cle}}}$}

\State \textbf{Alternated training stage:} 
\State \textbf{for} $e=1$ to $E_a$:
\StateIndent{Construct poisoned samples using ${{G}_{\phi }}$}
\StateIndent{Freeze ${{G}_{\phi }}$ and update ${{f}_{\theta }}$ with ${{\mathcal{L}}_{\text{model}}}$}
\StateIndent{Freeze ${{f}_{\theta }}$ and update ${{G}_{\phi }}$ with ${{\mathcal{L}}_{\text{tgr}}}$}
\StateIndent{Compute ${{\mathcal{L}}_{\text{atk}}}$ on poisoned samples} 
\StateIndent{Update poisoned-sample weights:}
\StateIndent{\hspace{0.5em} ${{\omega }_{e}}\leftarrow Norm({{\omega }_{e}}\cdot (1+{{\eta }_{b}}\cdot Norm({{\mathcal{L}}_{\text{atk}}})))$}

\end{algorithmic}
\end{algorithm}
\vspace{-0.3em}

\vspace{0.5em}
\begin{obsbox}
Module 3: Operation-Aware Loss
\end{obsbox}

The standard time series backdoor paradigm trains the forecasting model and the trigger generator using a clean loss and an attack loss. We extend this paradigm by incorporating operation-aware losses and physical constraints. We first introduce two prediction-related losses:

\vspace{-0.5em}
\begin{equation}
\label{eq:24}
{{\mathcal{L}}_{\text{cle}}}=MSE(\hat{\bm{Y}},\bm{Y})
\end{equation}

\vspace{-0.5em}
\begin{equation}
\label{eq:25}
{{\mathcal{L}}_{\text{atk}}}=MSE({{\hat{\bm{Y}}}^{\text{tar}}},{\bm{Y}^{\text{tar}}})+{{\lambda }_{\text{dis}}}{{\mathcal{L}}_{\text{dis}}}({{\hat{\bm{Y}}}^{\text{tar}}},{\bm{Y}^{\text{tar}}})
\end{equation}

The attack loss consists of two terms. The first term enforces point-wise matching between the triggered prediction and the target pattern. The second term measures their distributional discrepancy, since distributional shifts can also significantly affect power system operations~\cite{chen2026reshape}. Specifically, the maximum mean discrepancy (MMD) is adopted in ${\mathcal{L}}_{\text{dis}}$.

The training loss for the forecasting model is composed of the above two losses:

\vspace{-0.5em}
\begin{equation}
\label{eq:26}
{{\mathcal{L}}_{\text{model}}}={{\mathcal{L}}_{\text{cle}}}+\alpha {{\mathcal{L}}_{\text{atk}}}
\end{equation}

For the trigger generator, we further consider how deviations from the target affect downstream operation. Our motivation is that, when the forecasting model cannot fully learn the predefined target due to limited capacity, it should at least prioritize the error directions that are most harmful to downstream operations. To achieve this, we introduce a lightweight surrogate network ${{q}_{\varphi }}$, which estimates the degradation of attack effectiveness when the triggered prediction deviates from the target:

\vspace{-0.5em}
\begin{equation}
\label{eq:27}
({{\hat{\bm{Y}}}^{\text{tar}}}-{\bm{Y}^{\text{tar}}})\to \Delta {{\mathcal{J}}_{\text{op}}}
\end{equation}

The main challenge is obtaining training data for ${{q}_{\varphi }}$. In practice, we sample perturbations around the predefined target and compute the corresponding changes in the downstream objective. These perturbations are applied to the target amplitude and target pattern shape, such as adding or multiplying a constant factor, or distorting the local pattern shape.

The final training loss for the trigger generator is:

\vspace{-0.5em}
\begin{equation}
\label{eq:28}
{{\mathcal{L}}_{\text{tgr}}}={{\mathcal{L}}_{\text{cle}}}+{{\beta }_{1}}{{\mathcal{L}}_{\text{atk}}}+{{\beta }_{2}}{{q}_{\varphi }}({{\hat{\bm{Y}}}^{\text{tar}}},{\bm{Y}^{\text{tar}}})+{{\beta }_{3}}{{\mathcal{L}}_{\text{ste}}}
\end{equation}

\noindent
Here, ${\mathcal{L}}_{\text{ste}}$ denotes the stealthiness penalty, including smoothness constraints~\cite{xiang2025badtime}, physical range constraints, and other domain-specific restrictions.

\begin{table*}[t]
\centering
\caption{Main results. best results are in \textbf{bold}, and the second best is \underline{underlined}.}
\label{tab:main results}
\renewcommand{\arraystretch}{1.2}  
\setlength{\tabcolsep}{5pt}  
\newcommand{\patchgray}[1]{\textcolor{gray}{#1}}
\resizebox{\textwidth}{!}{
\begin{tabular}{c|c||ccc|ccc|ccc||ccc|ccc||ccc}
\toprule
\multicolumn{2}{c||}{\multirow{3}{*}{Methods}} & \multicolumn{9}{c||}{\includegraphics[height=1.0em]{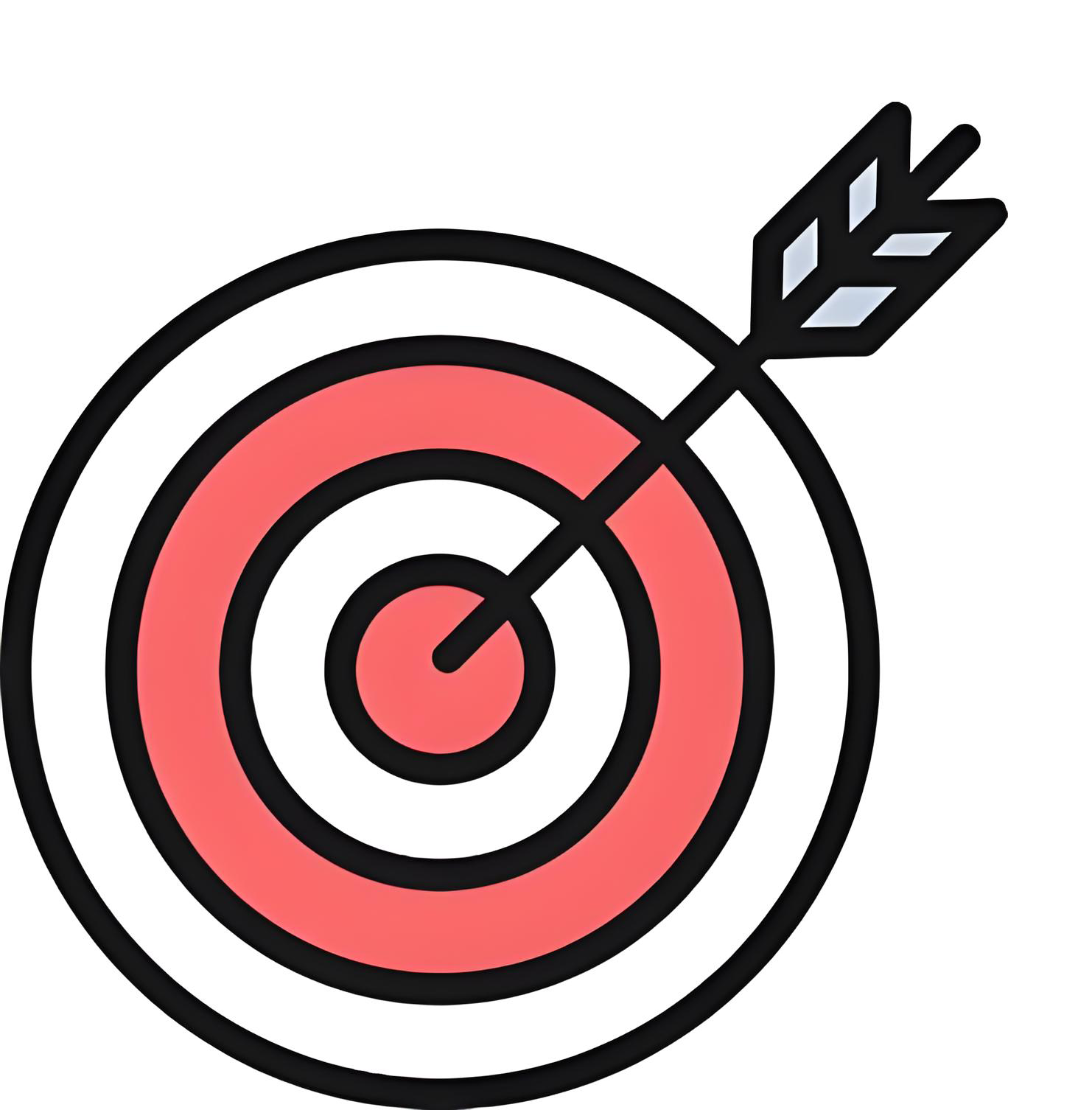}  \textit{\textbf{Backdoor Approaches}}} & \multicolumn{6}{c||}{\includegraphics[height=1.0em]{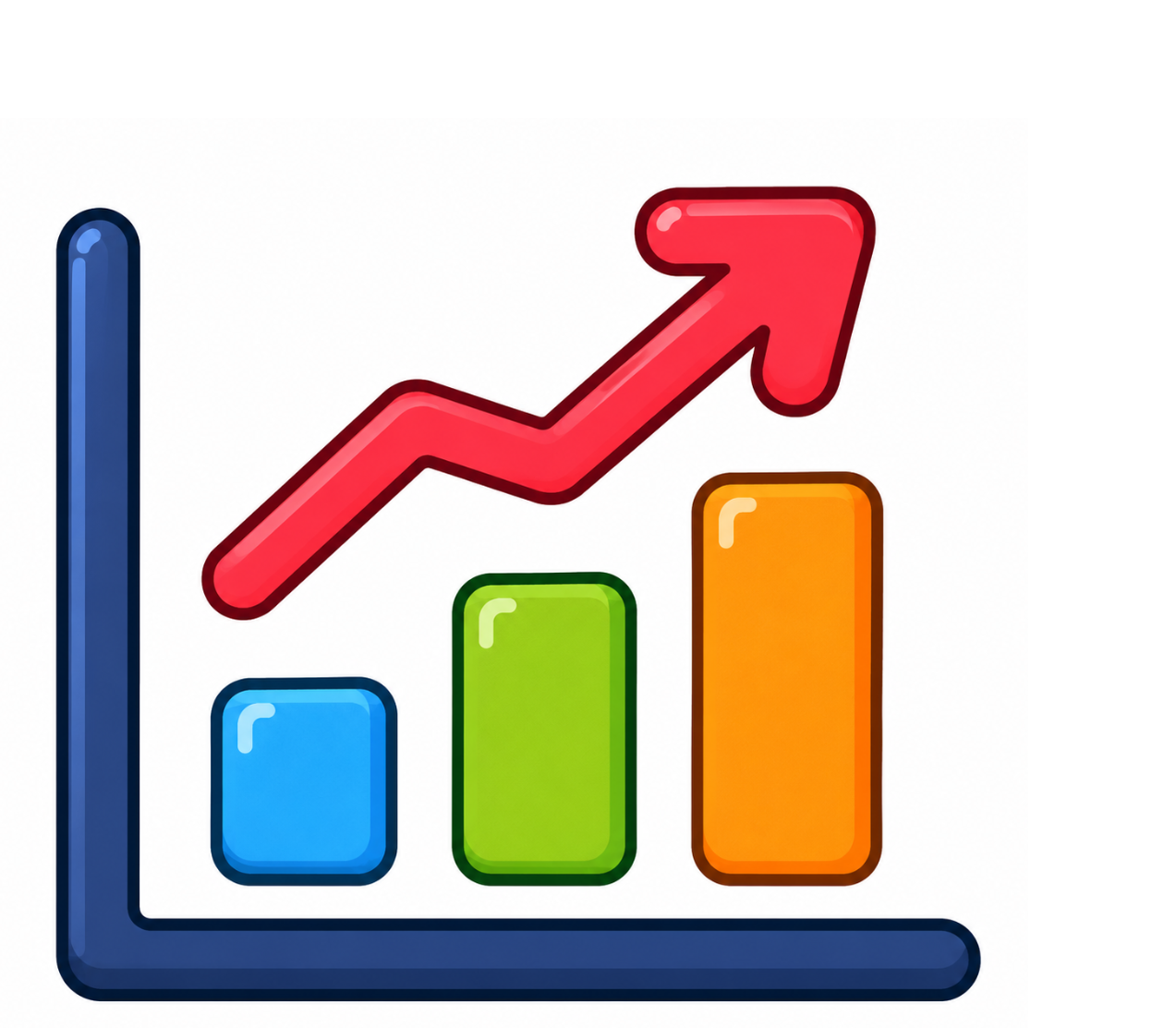}  \textit{\textbf{Traditional Cost-Oriented Approaches}}} & \multicolumn{3}{c}{\includegraphics[height=1.0em]{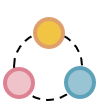}  \textit{\textbf{Pure Model}}} \\
\cline{3-20}
\multicolumn{2}{c||}{} & \multicolumn{3}{c|}{GridTroj} & \multicolumn{3}{c|}{BackTime} & \multicolumn{3}{c||}{BadTime} & \multicolumn{3}{c|}{CODP} & \multicolumn{3}{c||}{COAA} & \multicolumn{3}{c}{PatchTST}\\
\cline{3-20}
\multicolumn{2}{c||}{} & Atk. E & gap & viol(\%) & Atk. E & gap & viol(\%) & Atk. E & gap & viol(\%) & Atk. E & gap & viol(\%) & Atk. E & gap & viol(\%) & Atk. E & gap & viol(\%) \\
\midrule
\multirow{5}{*}{\rotatebox{90}{Task A}} 
& 13  & \textbf{5.473} & \textbf{1.028} & \textbf{0.989} & 3.05  & 0.580 & 0.191 & \underline{4.273} & \underline{0.782} & \underline{0.504} & 2.293 & 0.360 & 0.409 & 2.618 & 0.343 & 0.220 & $\backslash$ & \patchgray{0.135} & \patchgray{0.094} \\
& 33  & \textbf{7.467} & \textbf{1.232} & \textbf{1.141} & 3.830 & 0.655 & 0.777 & \underline{4.581} & \underline{0.893} & \underline{1.033} & 4.027 & 0.639 & 0.986 & 4.056 & 0.507 & 0.717 & $\backslash$ & \patchgray{0.167} & \patchgray{0.280} \\
& 69  & \textbf{4.784} & \textbf{0.776} & \textbf{0.727} & 2.02  & 0.349 & 0.550 & \underline{2.418} & \underline{0.399} & 0.589 & 1.270 & 0.188 & \underline{0.626} & 1.321 & 0.185 & 0.439 & $\backslash$ & \patchgray{0.082} & \patchgray{0.061} \\
& 123 & \textbf{2.948} & \textbf{0.454} & \textbf{0.322} & 1.627 & 0.262 & 0.195 & \underline{2.177} & \underline{0.320} & \underline{0.208} & 1.109 & 0.152 & 0.191 & 0.992 & 0.121 & 0.182 & $\backslash$ & \patchgray{0.056} & \patchgray{0.059} \\
\cline{2-20}
& Avg & \textbf{5.168} & \textbf{0.873} & \textbf{0.795} & 2.636 & 0.462 & 0.428 & \underline{3.362} & \underline{0.599} & \underline{0.584} & 2.175 & 0.335 & 0.553 & 2.247 & 0.289 & 0.390 & $\backslash$ & \patchgray{0.110} & \patchgray{0.124} \\

\midrule
\multirow{5}{*}{\rotatebox{90}{Task B}} 
& 13  & \textbf{1.196} & \textbf{0.189} & \textbf{0.301} & 0.901 & 0.144 & 0.137 & \underline{1.033} & \underline{0.157} & \underline{0.179} & 0.788 & 0.115 & 0.154 & 0.768 & 0.096 & 0.121 & $\backslash$ & \patchgray{0.010} & \patchgray{0.066} \\
& 33  & \textbf{1.297} & \textbf{0.240} & \textbf{0.372} & 1.018 & 0.167 & 0.160 & \underline{1.068} & \underline{0.188} & 0.208 & 1.047 & 0.180 & \underline{0.208} & 0.993 & 0.148 & 0.171 & $\backslash$ & \patchgray{0.018} & \patchgray{0.103} \\
& 69  & \textbf{0.975} & \textbf{0.157} & \textbf{0.219} & 0.805 & 0.109 & 0.171 & \underline{0.844} & \underline{0.130} & \underline{0.155} & 0.694 & 0.111 & 0.148 & 0.634 & 0.092 & 0.123 & $\backslash$ & \patchgray{0.014} & \patchgray{0.044} \\
& 123 & \textbf{0.885} & \textbf{0.154} & \textbf{0.220} & 0.584 & 0.111 & 0.146 & \underline{0.632} & \underline{0.124} & \underline{0.162} & 0.542 & 0.084 & 0.115 & 0.560 & 0.077 & 0.096 & $\backslash$ & \patchgray{0.009} & \patchgray{0.050} \\
\cline{2-20}
& Avg & \textbf{1.088} & \textbf{0.185} & \textbf{0.278} & 0.827 & 0.133 & 0.154 & \underline{0.894} & \underline{0.150} & \underline{0.176} & 0.768 & 0.123 & 0.155 & 0.739 & 0.103 & 0.128 & $\backslash$ & \patchgray{0.013} & \patchgray{0.066} \\

\midrule
\multirow{5}{*}{\rotatebox{90}{Task C}} 
& 13  & \textbf{2.197} & \textbf{0.290} & \textbf{0.488} & 1.413 & 0.185 & 0.235 & \underline{1.496} & \underline{0.212} & \underline{0.375} & 1.408 & 0.200 & 0.319 & 1.237 & 0.151 & 0.240 & $\backslash$ & \patchgray{0.028} & \patchgray{0.133} \\
& 33  & \textbf{2.098} & \textbf{0.403} & \textbf{0.873} & 1.129 & 0.254 & 0.558 & \underline{1.432} & \underline{0.295} & \underline{0.660} & 1.292 & 0.230 & 0.525 & 1.263 & 0.202 & 0.445 & $\backslash$ & \patchgray{0.052} & \patchgray{0.204} \\
& 69  & \textbf{1.276} & \textbf{0.222} & \underline{0.304} & 0.646 & 0.130 & 0.202 & \underline{0.821} & \underline{0.156} & \textbf{0.351} & 0.816 & 0.133 & 0.272 & 0.773 & 0.116 & 0.214 & $\backslash$ & \patchgray{0.024} & \patchgray{0.111} \\
& 123 & \textbf{1.037} & \textbf{0.196} & \textbf{0.265} & 0.612 & 0.120 & 0.184 & \underline{0.751} & \underline{0.133} & \underline{0.226} & 0.553 & 0.099 & 0.180 & 0.583 & 0.091 & 0.167 & $\backslash$ & \patchgray{0.021} & \patchgray{0.119} \\
\cline{2-20}
& Avg & \textbf{1.652} & \textbf{0.278} & \textbf{0.483} & 0.950 & 0.172 & 0.295 & \underline{1.125} & \underline{0.199} & \underline{0.403} & 1.017 & 0.166 & 0.324 & 0.964 & 0.140 & 0.267 & $\backslash$ & \patchgray{0.031} & \patchgray{0.142} \\

\bottomrule
\end{tabular}
}
\vspace{-0.3em}
\end{table*}

\section{Experiments}

In this section, we evaluate GridTroj on various downstream tasks and address three questions: 1) Do backdoor attacks pose a practical threat to downstream distribution network operation? 2) Can GridTroj achieve state-of-the-art attack performance compared to other baselines? 3) To what extent does the proposed task-aware design contribute to the attack effectiveness in power system scenarios?

\subsection{Experiment Setup}
\noindent
\textbf{Tasks.} Task A: Volt-Var Control, which minimizes power loss and voltage violation penalties. Task B: Economic Dispatch, which minimizes the total operating cost with energy storage, time-of-use prices and tie-line power exchange. Task C: Active/Reactive Coordinated Optimization, the most complex task, includes all devices in Tasks A and B and further incorporates flexible EV scheduling. Its objective is to minimize the total operating cost while considering voltage security and EV user satisfaction. We consider all combinations of the three downstream tasks and four standard distribution network topologies (IEEE-13, 33, 69, 123 bus systems) to evaluate the effectiveness and generalizability of GridTroj. We use a publicly available distribution network scenario dataset DDRE-33~\cite{chen2025large}. The detailed settings are provided in Appendix A. For each task, forecasting models first predict DER-related resources, based on which the optimization decision is solved. The obtained decision is then executed on the ground truth profiles, and the resulting objective variation is recorded.

\vspace{0.3em}
\noindent
\textbf{Baselines.} We compare GridTroj with compatible baselines, including: 1) standard backdoor attacks for TSF, BackTime~\cite{lin2024backtime} and BadTime~\cite{xiang2025badtime}; 2) cost-oriented data poisoning, CODP~\cite{chen2022vulnerability}; 3) adversarial attacks COAA~\cite{chen2022vulnerability}. In addition, we train a clean forecasting model to isolate the performance degradation introduced by attacks.

\vspace{0.3em}
\noindent
\textbf{Evaluation Metrics.} We evaluate each method using four metrics: 1) the relative objective gap against the ground truth optimal solution, denoted as \textit{gap} ↑; 2) the constraint violation rate, denoted as \textit{viol} ↑; 3) the forecasting error on clean inputs, measured by ${{MSE}_\text{c}}$ ↓; 4) the objective increase induced by a unit target deviation, referred to as Attack Efficiency, denoted as \textit{Atk. E} ↑. This metric measures whether an attack can induce larger downstream damage with smaller target manipulation.

\vspace{0.3em}
\noindent
\textbf{Implementation Details.} The dataset has a temporal resolution of 15 minutes. We consider three input lengths, 96, 288, and 672, corresponding to one, three, and seven days of historical observations, respectively. The prediction length is fixed to 96. Unless otherwise specified, we use PatchTST~\cite{nietime} as the forecasting backbone in the main experiments. We further evaluate GridTroj on other modern TSF models to demonstrate its model-agnostic property, as shown in Fig. 3. As reported in the original paper~\cite{chen2022vulnerability}, LR and LSTM are used as the forecasting backbones for CODP and COAA, respectively. The trigger generator adopts a two-layer GCN with a hidden dimension of 64. The surrogate network is a three-layer MLP with a hidden dimension of 32. 
During training, the forecasting model is first pretrained for 20 epochs, after which the trigger generator and the forecasting model are optimized alternately. We use the Adam optimizer with an initial learning rate of 0.0001. All experiments are conducted on Intel Xeon Gold 6254 processor, NVIDIA A800 GPU.

\begin{table}[t]
\centering
\caption{Forecasting performance under clean input}
\label{tab:forecasting_backbone}
\setlength{\tabcolsep}{6pt}
\newcommand{\patchgray}[1]{\textcolor{gray}{#1}}
\resizebox{\columnwidth}{!}{
\begin{tabular}{cc||cccc||c}
\toprule
\multicolumn{2}{c||}{\multirow{2}{*}{Methods}}
  & \multicolumn{4}{c||}{Attacks}
  & \multicolumn{1}{c}{Pure} \\
\cmidrule(lr){3-6} \cmidrule(lr){7-7}
\multicolumn{2}{c||}{}
  & GridTroj & BackTime & BadTime & CODP & PatchTST \\
\midrule
\multirow{2}{*}{13}
  & MAE\textsubscript{c}  & 0.051 & 0.052 & 0.048 & 0.066 & \patchgray{0.044} \\
  & MSE\textsubscript{c}  & 0.018 & 0.018 & 0.017 & 0.025 & \patchgray{0.017} \\
\midrule
\multirow{2}{*}{33}
  & MAE\textsubscript{c}  & 0.172 & 0.203 & 0.185 & 0.271 & \patchgray{0.163} \\
  & MSE\textsubscript{c}  & 0.050 & 0.066 & 0.052 & 0.114 & \patchgray{0.046} \\
\midrule
\multirow{2}{*}{69}
  & MAE\textsubscript{c}  & 0.146 & 0.170 & 0.157 & 0.250 & \patchgray{0.139} \\
  & MSE\textsubscript{c}  & 0.047 & 0.059 & 0.049 & 0.099 & \patchgray{0.041} \\
\midrule
\multirow{2}{*}{123}
  & MAE\textsubscript{c}  & 0.142 & 0.163 & 0.144 & 0.243 & \patchgray{0.137} \\
  & MSE\textsubscript{c}  & 0.044 & 0.057 & 0.045 & 0.096 & \patchgray{0.042} \\
\bottomrule
\end{tabular}
}
\vspace{-0.3em}
\end{table}

\begin{figure}[!t]
\centering
\includegraphics[width=0.95\columnwidth]{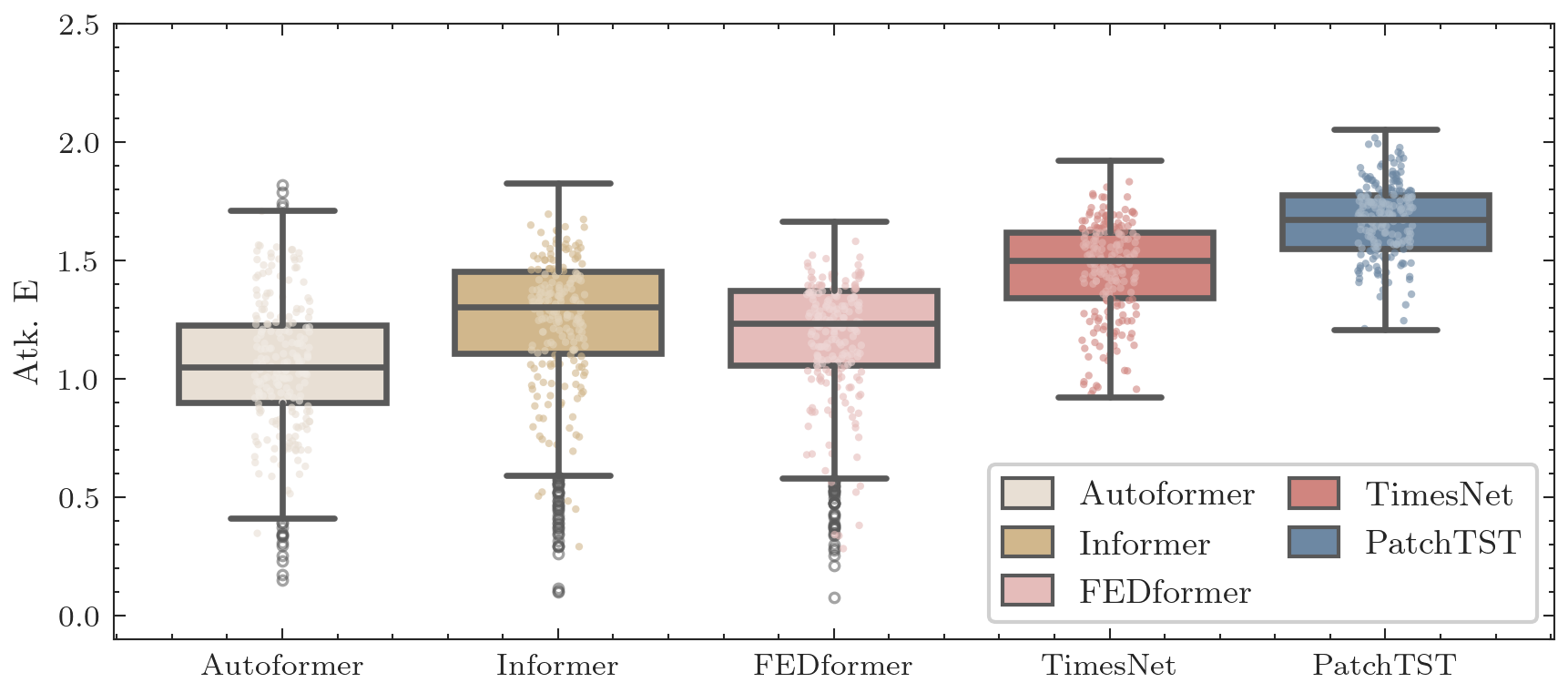}
\caption{Atk.E matrix on different forecasting backbone}
\label{fig:x}
\vspace{-0.7em}
\end{figure}

\subsection{Main Results}
Table I reports the performance of different attack methods.

1) \textbf{GridTroj consistently outperforms all baselines.} Among the standard backdoor methods, BadTime achieves a higher objective gap than BackTime by optimizing the trigger position for a higher attack success rate. However, both yield similar Atk. E because they focus solely on minimizing forecasting errors without task awareness. In contrast, GridTroj significantly boosts Atk. E through its intention-aware strategy, inflicting severe operational damage while maintaining low target deviation.

2) \textbf{Backdoor attacks generally surpass adversarial counterparts.} Although the adversarial baseline COAA utilizes a cost-oriented strategy, its objective gap remains lower than CODP's. In addition, both CODP and COAA yield larger MSE\textsubscript{c} than Transformer-based methods, confirming the forecasting advantage of modern deep forecasting models over traditional models.

3) \textbf{Operational security hazards:} The \textit{viol} metric denotes the percentage of actual voltage violations outside the [0.95, 1.05] pu range during real-world operations driven by forecast-based decisions. For Task C, EV energy shortages and charging interruption counts are also evaluated. GridTroj induces the highest \textit{viol} due to its downstream-disruptive target design. CODP also demonstrates high \textit{viol} but suffers from a much larger variance, which we attribute to its inherently poor forecasting capability.

Table II evaluates forecasting errors under clean inputs to assess stealthiness. A successful backdoor must remain hidden during normal operations. GridTroj achieves the MSE\textsubscript{c} closest to the benign model, mainly due to its poison sample selection strategy, which establishes the backdoor with minimal sample manipulation.

As a generalized framework, GridTroj accommodates various forecasting backbones as plug-and-play modules. Fig. 3 plots the boxplots of Atk. E on Task C across different backbones. PatchTST yields the highest and most stable Atk. E, while Autoformer exhibit larger variances. Crucially, this performance hierarchy mirrors the rankings of these models in the time series domain, demonstrating that a more potent forecasting backbone directly amplifies GridTroj's efficacy.

\subsection{Case Study}
This section presents a comprehensive case study on Task C with the IEEE-33 bus system to illustrate the full backdoor attack–forecasting–optimization chain. As shown in Fig. 4, the attacker first poisons the training data to obtain a backdoored model and subsequently injects a well-crafted trigger to elicit the attacker-specified target output.

\begin{figure}[h]
\centering
\includegraphics[width=\columnwidth]{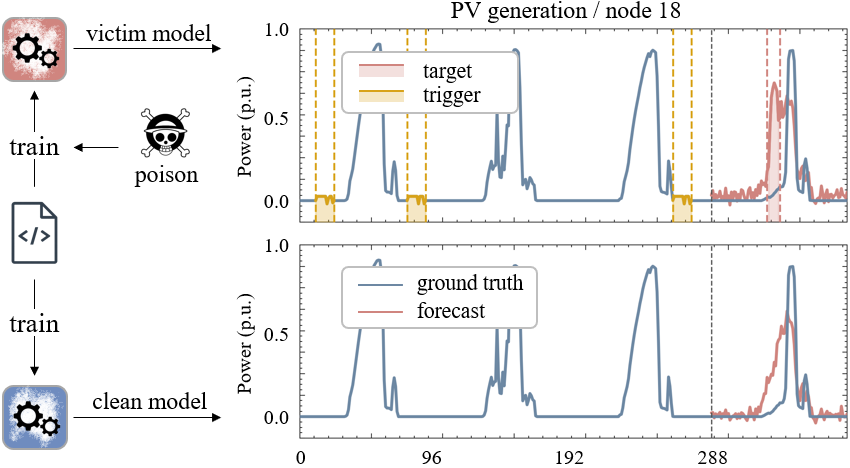}
\caption{Attacked forecasting and clean forecasting}
\label{fig:x}
\vspace{-0.5em}
\end{figure}

We then solve the downstream operation problem with a commercial solver and examine how the malicious forecast changes operational decisions. We first focus on two discrete voltage-control variables: the OLTC tap position and capacitor bank steps (SC). Fortunately, these two decisions remain unaffected, as shown in Fig. 5. The OLTC tap position stays at or near the highest level for most of the day, while SC provides more reactive compensation around noon to reduce active power losses and mitigate voltage violations.

\begin{figure}[h]
\centering
\includegraphics[width=0.95\columnwidth]{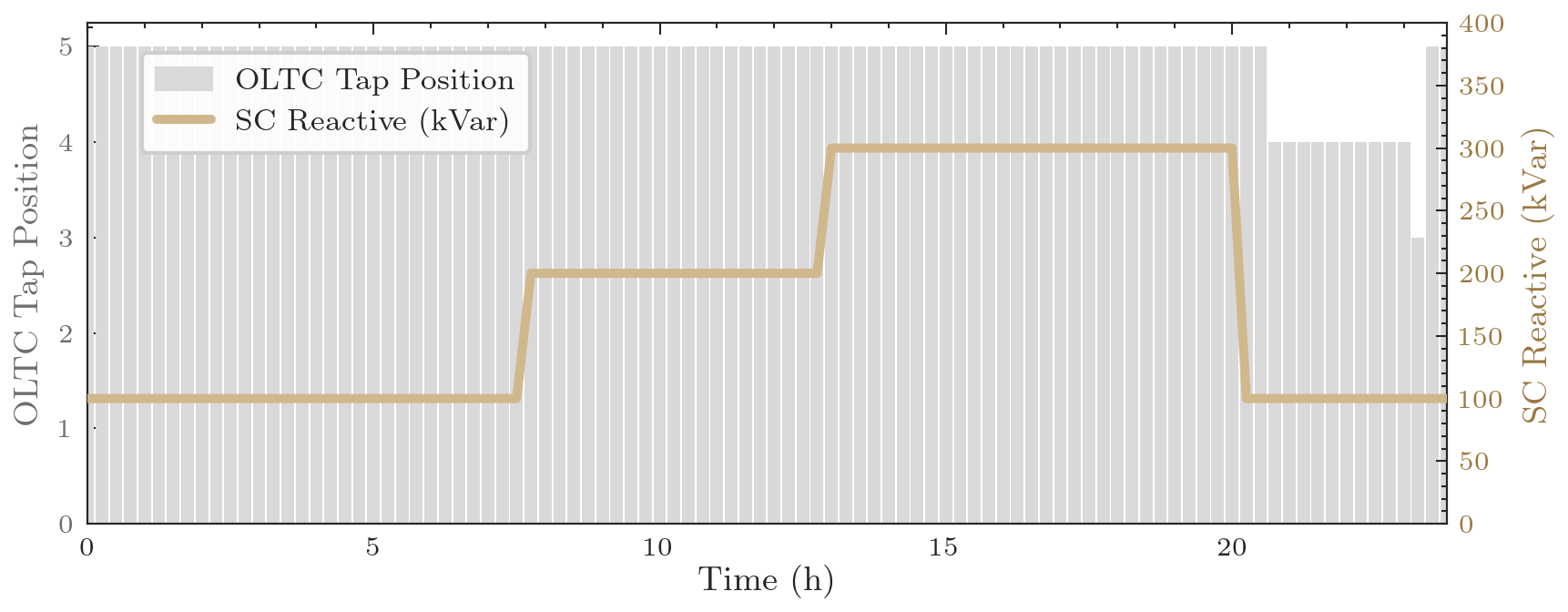}
\caption{OLTC tap position and SC power}
\label{fig:x}
\vspace{-0.5em}
\end{figure}

However, the ESS and EV schedules are severely compromised. As shown in Fig. 6, the overestimated PV forecast leads to massive EV charging curtailment and interruption under the evening fast-charging scenario (\ding{174}). Meanwhile, the ESS fails to pre-charge sufficient energy, leading to a supply deficit (\ding{172}), while the distribution network is forced to purchase power during peak-price periods (\ding{173}).

\begin{figure}[h]
\centering
\includegraphics[width=\columnwidth]{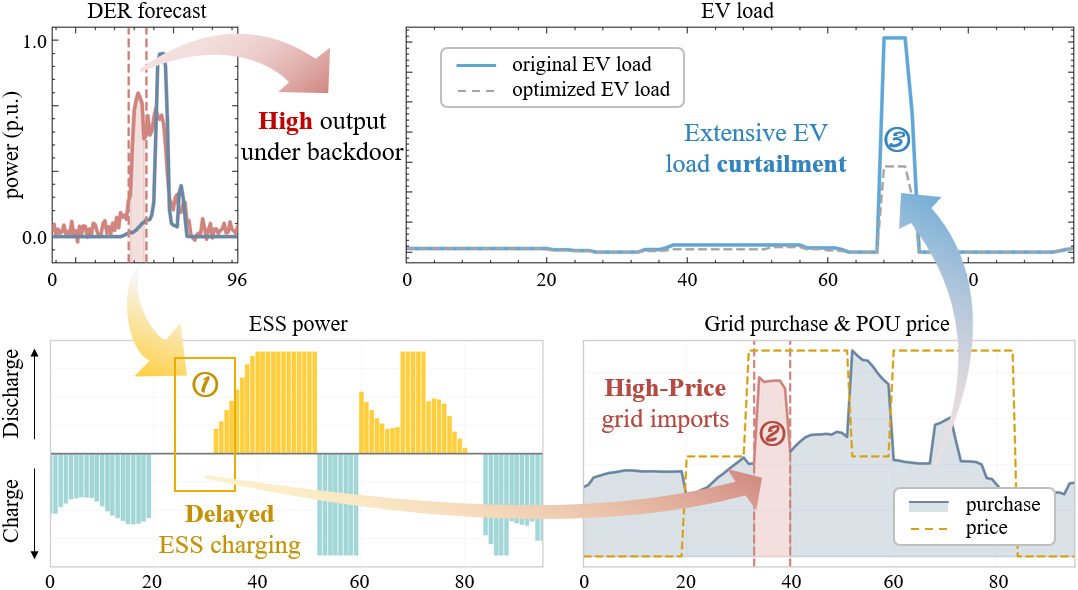}
\caption{ESS, EV load and grid purchased power}
\label{fig:x}
\vspace{-0.5em}
\end{figure}

Here we further discuss two operational scenarios based on grid capabilities. 1) Normal Case. The grid can dynamically restrict EV charging power at any time, which merely incurs extra curtailment penalties and escalates operational costs. 2) Severe Case. The grid cannot actively curtail EV loads in real time, causing a surge in concentrated fast-charging that overwhelms the network and triggers severe security failures (the solver returns \textit{infeasible} under this setting). Note that the exact outcome depends on specific utility configurations, which is beyond the scope of this paper. Here, we focus on validating the potential backdoor threat.

\subsection{Ablation and Discussion}

Table III presents the ablation analysis of GridTroj's core components. Removing the \textit{Intention Planner} reduces the framework to the task-agnostic target selection of BackTime, causing a sharp decline in both gap and Atk. E. Among individual designs, Designs 1 and 2 degrade Atk. E most severely, proving the importance of task-aware intention. Furthermore, stripping the \textit{Backdoor Realizer} significantly cripples backdoor capabilities, which not only lowers the operational gap but also elevates the target fitting error MAE\textsubscript{t}.

\begin{table}[t]
\vspace{-0.6em}
\centering
\caption{Ablation study on model architecture}
\renewcommand{\arraystretch}{1.0}  
\setlength{\tabcolsep}{5pt}
\resizebox{\columnwidth}{!}{
\begin{tabular}{l|cc|cc|cc}
\toprule
& \multicolumn{2}{c|}{Task A} & \multicolumn{2}{c|}{Task B} & \multicolumn{2}{c}{Task C} \\
\cline{2-7}

& Atk. E & gap & Atk. E & gap & Atk. E & gap \\
\midrule

GridTroj & \textbf{5.168} & \textbf{0.873} & \textbf{1.088} & \textbf{0.185} & \textbf{1.652} & \textbf{0.278} \\
\midrule
w/o Intention Planner & 3.085 & 0.522 & 0.848 & 0.136 & 1.030 & 0.177 \\
$\rightarrow$ w/o design 1 & 3.904 & 0.677 & 0.884 & 0.152 & 1.306 & 0.219 \\
$\rightarrow$ w/o design 2 & 4.311 & 0.740 & 0.963 & 0.165 & 1.378 & 0.231 \\
$\rightarrow$ w/o design 3 & 5.066 & 0.829 & 1.024 & 0.170 & 1.501 & 0.240 \\
$\rightarrow$ w/o design 4 & 4.505 & 0.727 & 1.010 & 0.166 & 1.492 & 0.226 \\
\midrule
w/o Backdoor Realizer & 4.281 & 0.694 & 0.913 & 0.148 & 1.442 & 0.229 \\
$\rightarrow$ w/o module 1 & 4.771 & 0.762 & 1.020 & 0.173 & 1.564 & 0.258 \\
$\rightarrow$ w/o module 3 & 3.969 & 0.705 & 0.898 & 0.164 & 1.387 & 0.255 \\
\bottomrule
\end{tabular}
}
\vspace{-0.9em}
\end{table}

\noindent
\textbf{Stealthiness Comparison.} To further evaluate the stealthiness of GridTroj, we adopt two time series anomaly detectors, USAD~\cite{audibert2020usad} and FCVAE~\cite{wang2024revisiting}. Both detectors are trained on clean data and tested on poisoned data. As reported in Table IV, GridTroj yields the lowest F1-score and AUC across the board. Its AUC approaches 0.5, indicating that the detectors are nearly reduced to random guessing. 
We also conduct physical-constraint checks. The results show that GridTroj produces almost no obvious physical violations, while BackTime and BadTime may generate detectable abnormal patterns, such as wind speed-power mismatch or positive PV output during nighttime. This verifies the stealthiness of our attack.

\begin{table}[h]
\vspace{-0.6em}
\centering
\caption{Anomaly detection results}
\renewcommand{\arraystretch}{1.0}  
\setlength{\tabcolsep}{5pt}
\resizebox{\columnwidth}{!}{
\begin{tabular}{l|cc|cc|cc|cc}
\toprule
& \multicolumn{2}{c|}{GridTroj} & \multicolumn{2}{c|}{BackTime} & \multicolumn{2}{c}{BadTime} & \multicolumn{2}{c}{CODP}\\
\cline{2-9}

& F1 & AUC & F1 & AUC & F1 & AUC & F1 & AUC \\
\midrule

USAD & \textbf{0.019} & \textbf{0.504} & 0.413 & 0.660 & 0.327 & 0.523 & 0.679 & 0.782 \\
FCVAE & \textbf{0.019} & \textbf{0.505} & 0.498 & 0.728 & 0.379 & 0.606 & 0.744 & 0.821 \\
\bottomrule
\end{tabular}
}
\vspace{-0.6em}
\end{table}

\noindent
\textbf{Discussion on Target Optimization.} In the \textit{Intention Planner}, GridTroj designs targets to increase downstream operational damage. The min-max formulations in Section IV provide an operation-guided principle for target construction. In implementation, similar to~\cite{chen2022vulnerability}, we instantiate this principle through controllable hyperparameter search rather than claiming exact global optimality. This is reasonable because the final attack effect is also constrained by backdoor capability. Even if a mathematically optimal target were available, the model may not perfectly reproduce it after trigger activation. Therefore, we treat target construction as a practical, tunable design. Additional sensitivity analysis is provided in Appendix C.

\section{Conclusion}

In this paper, we reveal the backdoor threat in forecast-based distribution network operation and propose GridTroj, a unified backdoor framework tailored to this scenario. By centering the attack design on the attacker’s operational intention, GridTroj can effectively translate malicious forecasting targets into downstream operational disruptions. Extensive experiments demonstrate the effectiveness and stealthiness of the proposed framework. We hope this work will motivate further research on backdoor defense for secure power system operation.

\section*{Acknowledgments}

This work was supported by National Science and Technology Major Project of the Ministry of Science and Technology of China (No.2025ZD0805900) and National Natural Science Foundation of China (U24B6010)



 

\bibliographystyle{IEEEtran}
\bibliography{References}

@article{graves2012long,
  title={Long short-term memory},
  author={Graves, Alex},
  journal={Supervised sequence labelling with recurrent neural networks},
  pages={37--45},
  year={2012},
  publisher={Springer}
}

@inproceedings{zhou2021informer,
  title={Informer: Beyond efficient transformer for long sequence time-series forecasting},
  author={Zhou, Haoyi and Zhang, Shanghang and Peng, Jieqi and Zhang, Shuai and Li, Jianxin and Xiong, Hui and Zhang, Wancai},
  booktitle={Proceedings of the AAAI conference on artificial intelligence},
  volume={35},
  number={12},
  pages={11106--11115},
  year={2021}
}

@article{wu2021autoformer,
  title={Autoformer: Decomposition transformers with auto-correlation for long-term series forecasting},
  author={Wu, Haixu and Xu, Jiehui and Wang, Jianmin and Long, Mingsheng},
  journal={Advances in neural information processing systems},
  volume={34},
  pages={22419--22430},
  year={2021}
}

@inproceedings{zhou2022fedformer,
  title={Fedformer: Frequency enhanced decomposed transformer for long-term series forecasting},
  author={Zhou, Tian and Ma, Ziqing and Wen, Qingsong and Wang, Xue and Sun, Liang and Jin, Rong},
  booktitle={International conference on machine learning},
  pages={27268--27286},
  year={2022},
  organization={PMLR}
}

@inproceedings{nietime,
  title={A Time Series is Worth 64 Words: Long-term Forecasting with Transformers},
  author={Nie, Yuqi and Nguyen, Nam H and Sinthong, Phanwadee and Kalagnanam, Jayant},
  booktitle={The Eleventh International Conference on Learning Representations}
}

@article{chen2025large,
  title={A Large-Scale Dataset of Distributed Renewable Energy Scenarios on the IEEE-33 Bus Network},
  author={Chen, Yuxuan and Xie, Haipeng and Huang, Wenqi and Li, Peng},
  journal={Scientific Data},
  year={2025},
  publisher={Nature Publishing Group UK London}
}

@article{lin2024backtime,
  title={Backtime: Backdoor attacks on multivariate time series forecasting},
  author={Lin, Xiao and Liu, Zhining and Fu, Dongqi and Qiu, Ruizhong and Tong, Hanghang},
  journal={Advances in Neural Information Processing Systems},
  volume={37},
  pages={131344--131368},
  year={2024}
}

@article{xiang2025badtime,
  title={BadTime: An Effective Backdoor Attack on Multivariate Long-Term Time Series Forecasting},
  author={Xiang, Kunlan and Yang, Haomiao and Hao, Meng and Jiang, Wenbo and Wang, Haoxin and Huang, Shiyue and Li, Shaofeng and Liu, Yijing and Guo, Ji and Niyato, Dusit},
  journal={arXiv preprint arXiv:2508.04189},
  year={2025}
}

@article{chen2022vulnerability,
  title={Vulnerability and impact of machine learning-based inertia forecasting under cost-oriented data integrity attack},
  author={Chen, Yan and Sun, Mingyang and Chu, Zhongda and Camal, Simon and Kariniotakis, George and Teng, Fei},
  journal={IEEE Transactions on Smart Grid},
  volume={14},
  number={3},
  pages={2275--2287},
  year={2022},
  publisher={IEEE}
}

@inproceedings{huynh2024combat,
  title={Combat: Alternated training for effective clean-label backdoor attacks},
  author={Huynh, Tran and Nguyen, Dang and Pham, Tung and Tran, Anh},
  booktitle={Proceedings of the AAAI Conference on Artificial Intelligence},
  volume={38},
  number={3},
  pages={2436--2444},
  year={2024}
}

@article{chen2026reshape,
  title={Reshape: Adversarial Attack on Probabilistic Wind Power Forecasting},
  author={Chen, Yan and Chen, Boli and Sun, Mingyang},
  journal={IEEE Transactions on Smart Grid},
  year={2026},
  publisher={IEEE}
}

@inproceedings{chen2019exploiting,
  title={Exploiting vulnerabilities of load forecasting through adversarial attacks},
  author={Chen, Yize and Tan, Yushi and Zhang, Baosen},
  booktitle={Proceedings of the tenth ACM international conference on future energy systems},
  pages={1--11},
  year={2019}
}

@article{liu2025robust,
  title={Robust photovoltaic power forecasting against multi-modal adversarial attack via deep reinforcement learning},
  author={Liu, Jingxuan and Zang, Haixiang and Cheng, Lilin and Ding, Tao and Wei, Zhinong and Sun, Guoqiang},
  journal={IEEE Transactions on Sustainable Energy},
  year={2025},
  publisher={IEEE}
}

@article{chao2025cyber,
  title={Cyber Resilience of Three-phase Unbalanced Distribution System Restoration under Sparse Adversarial Attack on Load Forecasting},
  author={Chao, Chen and Ma, Zixiao and Zhang, Ziang},
  journal={arXiv preprint arXiv:2510.03635},
  year={2025}
}

@article{kingma2013auto,
  title={Auto-encoding variational bayes},
  author={Kingma, Diederik P and Welling, Max},
  journal={arXiv preprint arXiv:1312.6114},
  year={2013}
}

@article{gu2017badnets,
  title={Badnets: Identifying vulnerabilities in the machine learning model supply chain},
  author={Gu, Tianyu and Dolan-Gavitt, Brendan and Garg, Siddharth},
  journal={arXiv preprint arXiv:1708.06733},
  year={2017}
}

@inproceedings{liu2018trojaning,
  title={Trojaning attack on neural networks},
  author={Liu, Yingqi and Ma, Shiqing and Aafer, Yousra and Lee, Wen-Chuan and Zhai, Juan and Wang, Weihang and Zhang, Xiangyu},
  booktitle={25th Annual Network And Distributed System Security Symposium (NDSS 2018)},
  year={2018},
  organization={Internet Soc}
}

@article{wang2020backdoor,
  title={Backdoor attacks against transfer learning with pre-trained deep learning models},
  author={Wang, Shuo and Nepal, Surya and Rudolph, Carsten and Grobler, Marthie and Chen, Shangyu and Chen, Tianle},
  journal={IEEE Transactions on Services Computing},
  volume={15},
  number={3},
  pages={1526--1539},
  year={2020},
  publisher={IEEE}
}

@inproceedings{ning2022trojanflow,
  title={Trojanflow: A neural backdoor attack to deep learning-based network traffic classifiers},
  author={Ning, Rui and Xin, Chunsheng and Wu, Hongyi},
  booktitle={IEEE INFOCOM 2022-IEEE Conference on Computer Communications},
  pages={1429--1438},
  year={2022},
  organization={IEEE}
}

@inproceedings{ding2022towards,
  title={Towards backdoor attack on deep learning based time series classification},
  author={Ding, Daizong and Zhang, Mi and Huang, Yuanmin and Pan, Xudong and Feng, Fuli and Jiang, Erling and Yang, Min},
  booktitle={2022 IEEE 38th International Conference on Data Engineering (ICDE)},
  pages={1274--1287},
  year={2022},
  organization={IEEE}
}

@inproceedings{huang2025revisiting,
  title={Revisiting backdoor attacks on time series classification in the frequency domain},
  author={Huang, Yuanmin and Zhang, Mi and Wang, Zhaoxiang and Li, Wenxuan and Yang, Min},
  booktitle={Proceedings of the ACM on Web Conference 2025},
  pages={1795--1810},
  year={2025}
}

@incollection{box2013box,
  title={Box and Jenkins: time series analysis, forecasting and control},
  author={Box, George},
  booktitle={A Very British Affair: Six Britons and the Development of Time Series Analysis During the 20th Century},
  pages={161--215},
  year={2013},
  publisher={Springer}
}

@article{hyndman2002state,
  title={A state space framework for automatic forecasting using exponential smoothing methods},
  author={Hyndman, Rob J and Koehler, Anne B and Snyder, Ralph D and Grose, Simone},
  journal={International Journal of forecasting},
  volume={18},
  number={3},
  pages={439--454},
  year={2002},
  publisher={Elsevier}
}

@inproceedings{wang2024revisiting,
  title={Revisiting vae for unsupervised time series anomaly detection: A frequency perspective},
  author={Wang, Zexin and Pei, Changhua and Ma, Minghua and Wang, Xin and Li, Zhihan and Pei, Dan and Rajmohan, Saravan and Zhang, Dongmei and Lin, Qingwei and Zhang, Haiming and others},
  booktitle={Proceedings of the ACM web conference 2024},
  pages={3096--3105},
  year={2024}
}

@inproceedings{audibert2020usad,
  title={Usad: Unsupervised anomaly detection on multivariate time series},
  author={Audibert, Julien and Michiardi, Pietro and Guyard, Fr{\'e}d{\'e}ric and Marti, S{\'e}bastien and Zuluaga, Maria A},
  booktitle={Proceedings of the 26th ACM SIGKDD international conference on knowledge discovery \& data mining},
  pages={3395--3404},
  year={2020}
}

@article{zografopoulos2023distributed,
  title={Distributed energy resources cybersecurity outlook: Vulnerabilities, attacks, impacts, and mitigations},
  author={Zografopoulos, Ioannis and Hatziargyriou, Nikos D and Konstantinou, Charalambos},
  journal={IEEE Systems Journal},
  volume={17},
  number={4},
  pages={6695--6709},
  year={2023},
  publisher={IEEE}
}


 




\vspace{-0.5em}
\section*{Appendix}

\subsection{Optimization Tasks Formulations}

\renewcommand{\thefigure}{A\arabic{figure}}
\setcounter{figure}{0}
\renewcommand{\theequation}{A\arabic{equation}}
\setcounter{equation}{0}
\renewcommand{\thetable}{A\arabic{table}}
\setcounter{table}{0}

\textit{Due to space limitations, a complete variable nomenclature table
will be provided in the second-round revision.} Results are averaged over 200 scenarios per task-feeder combination.

\noindent
\textbf{1) Network topology and DER placement}

We conduct experiments on four IEEE radial feeders:

\begin{table}[h] 
\vspace{-0.3em}
    \centering 
    \caption{Network configuration}
    \label{tab:dataset} 
    \renewcommand{\arraystretch}{0.9}  
    \setlength{\tabcolsep}{5pt} 
    \resizebox{\columnwidth}{!}{ 
    \begin{tabular}{ c| c c c c}
        \toprule
        \textbf{Device} & \textbf{IEEE-13} & \textbf{IEEE-33} & \textbf{IEEE-69} & \textbf{IEEE-123} \\ 
        \midrule
        PV & 634, 580, 684 & 18, 33 & 27, 50, 62 & 32, 50, 83, 88, 110 \\
        WT & / & 22, 25 & 18, 65 & 41, 57, 71 \\
        ESS & 671 & 10, 23, 31 & 12, 50 & 50, 88 \\
        EVs & 692 & 15, 21, 29 & 28, 51, 66 & 48, 65, 76, 96, 114 \\
        SC & 611, 675 & 16, 20, 31 & 12, 50, 62 & 65, 83, 88 \\
        SVC & 671, 675 & 11, 29 & 18, 51 & 50, 71, 110 \\
        \midrule
        Tie lines & 0 & 3 & 5 & 3 \\
        \bottomrule
    \end{tabular}
    }
    \vspace{-0.6em}
\end{table}

\noindent
\textbf{2) Task A — Volt-VAR Control}

Task A studies distribution-level VVC, which minimizes active loss and penalizes voltage violations:

\vspace{-0.8em}
\begin{equation}\label{eq}
\min;\sum_{\tau\in\Omega_\tau}\sum_{(u,v)\in\Omega_e} R_{uv},L_{uv,\tau}
+\lambda_v\sum_{\tau,n}\bigl(s_{n,\tau}^{+}+s_{n,\tau}^{-}\bigr)
\end{equation}

\noindent
subject to the key device constraints:

\vspace{-0.8em}
\begin{equation}\label{eq}
\sum\nolimits_{\tau=0}^{H-2} a_{\tau}^{\text{tap}}\le N^{\text{tap}},\qquad
\sum\nolimits_{\tau=0}^{H-2} a_{\tau,b}^{\text{cap}}\le N^{\text{cap}}
\end{equation}

\vspace{-0.8em}
\begin{equation}\label{eq}
q_{\tau,b}^{\text{cap}} = k_{\tau,b}^{\text{cap}} q^{\text{unit}}, \qquad
0 \le k_{\tau,b}^{\text{cap}} \le K^{\text{cap}}
\end{equation}


Eq. (A2)-(A3) bounds the daily actions of OLTC and models the discrete SC output.

\noindent
\textbf{3) Task B — Economic Dispatch}

Task B minimizes daily operating cost with grid purchase, ESS operation, PV curtailment, and tie-switch reconfiguration:

\vspace{-0.8em}
\begin{equation}\label{eq}
\resizebox{1.05\hsize}{!}{
$\min;
\sum_{\tau} \Bigl[
c_\tau^{\text{grid}} p_\tau^{\text{grid}}
+c^{\text{ess}}\sum_{k}\bigl(p_{\tau,k}^{\text{cha}}+p_{\tau,k}^{\text{dis}}\bigr)
+c^{\text{cur}}\sum_{n}p_{\tau,n}^{\text{cur}}
\Bigr]\Delta \tau
+c^{\text{sw}}\sum_{\tau,l} u_{\tau,l}^{\text{sw}}$
}
\end{equation}

\noindent
with the following representative constraints:

\vspace{-0.8em}
\begin{equation}\label{eq}
B_{\tau,k}=B_{\tau-1,k}
+\eta^{\text{cha}}p_{\tau,k}^{\text{cha}}\Delta \tau
-\frac{p_{\tau,k}^{\text{dis}}}{\eta^{\text{dis}}}\Delta \tau
\end{equation}

\vspace{-0.8em}
\begin{equation}\label{eq}
p_{\tau,k}^{\text{cha}}\le P_k^{\text{cha}}(1-z_{\tau,k}),\qquad
p_{\tau,k}^{\text{dis}}\le P_k^{\text{dis}}z_{\tau,k}
\end{equation}

\vspace{-0.8em}
\begin{equation}\label{eq}
|p_{\tau,l}^{\text{tie}}|\le M\omega_{\tau,l},\space
|q_{\tau,l}^{\text{tie}}|\le M\omega_{\tau,l},\space
L_{\tau,l}^{\text{tie}}\le M\omega_{\tau,l}
\end{equation}

Eqs. (A5)-(A6) describe ESS energy evolution and charge/discharge exclusiveness through binary mode $z_{\tau,k}$. The terminal SOC is restricted within 10\% of the initial value. Eq. (A7) gives the Big-M constraints for tie lines.

\noindent
\textbf{4) Task C — Active/Reactive Coordinated Optimization}

Task C combines the devices in Tasks A and B and further considers flexible EV charging. Let $\mathcal{C}_{\text{ED}}$ denote the objective in (A4). The objective is extended as:

\vspace{-0.8em}
\begin{equation}\label{eq}
\min;
\mathcal{C}_{\text{ED}}
+\lambda_v\sum{\tau,n}\bigl(s_{n,\tau}^{+}+s_{n,\tau}^{-}\bigr)
+\Pi^{\text{ev}}
\end{equation}

\noindent
The EV-specific constraints are:

\vspace{-0.8em}
\begin{equation}\label{eq}
p_{\tau,c}^{\text{ev}}=\bar{p}{\tau,c}^{\text{ev}}(1-\gamma{\tau,c}),\qquad
\gamma_{\tau,c}=\gamma_{\tau,c}^{1}+\gamma_{\tau,c}^{2}+\gamma_{\tau,c}^{3}
\end{equation}

\vspace{-0.8em}
\begin{equation}\label{eq}
\sum \bar{p}{\tau,c}^{\text{ev}}
\bigl(
d_1^{\text{ev}}\gamma{\tau,c}^{1}
+d_2^{\text{ev}}\gamma_{\tau,c}^{2}
+d_3^{\text{ev}}\gamma_{\tau,c}^{3}
\bigr)\Delta \tau
\end{equation}

\vspace{-0.8em}
\begin{equation}\label{eq}
B_{\tau,c}^{\text{ev}}=B_{\tau-1,c}^{\text{ev}}
+\eta^{\text{ev}}p_{\tau,c}^{\text{ev}}\Delta \tau,\qquad
\epsilon_c\ge B_c^{\text{req}}-B_{H,c}^{\text{ev}}
\end{equation}

\vspace{-0.8em}
\begin{equation}\label{eq}
\iota_{\tau,c}\ge v_{\tau-1,c}-v_{\tau,c},\qquad \forall,\tau\ge2
\end{equation}

\vspace{-0.8em}
\begin{equation}\label{eq}
\Pi_{\text{cur}}^{\text{ev}}
+d^{\text{short}}\sum_c\epsilon_c
+d^{\text{int}}\sum_{\tau,c}\iota_{\tau,c}
\end{equation}

Eq. (A10) defines EV charging after curtailment, where $\gamma_{\tau,c}$ is split into three tiers with $d_1^{\text{ev}}<d_2^{\text{ev}}<d_3^{\text{ev}}$. Eq. (A12) tracks EV charging energy and shortage slack. Eq. (A13) records charging interruptions.

The forecast-fixed decisions include all day-ahead discrete variables and the EV scheduling plan, since 15min resolution V2G dispatch control is difficult to implement in practice.

\subsection{Practical Feasibility and Threat Scope}

\renewcommand{\thefigure}{B\arabic{figure}}
\setcounter{figure}{0}
\renewcommand{\theequation}{B\arabic{equation}}
\setcounter{equation}{0}
\renewcommand{\thetable}{B\arabic{table}}
\setcounter{table}{0}

High-penetration DERs are increasingly operated by consumers and third-party vendors. Their heterogeneous architectures, communication dependencies, and protocol-level vulnerabilities expose DER data flows to malicious manipulation via compromised devices or pipelines~\cite{zografopoulos2023distributed}. 

Given this distributed nature, we assume each DER data stream or forecasting model is managed by an independent vendor with no cross-entity data access. Multi-attacker collaborative or game-theoretic scenarios are beyond the scope of this paper and left for future work.

\subsection{More Ablation and Hyperparameter Sensitivity Analysis}

\renewcommand{\thefigure}{C\arabic{figure}}
\setcounter{figure}{0}
\renewcommand{\theequation}{C\arabic{equation}}
\setcounter{equation}{0}
\renewcommand{\thetable}{C\arabic{table}}
\setcounter{table}{0}

\noindent
\textbf{Number of Poison Variables} $K$. This hyperparameter affects the model's ability to fit the target, directly influencing downstream operational disruption. The datasets comprise seven feature variables in total. We evaluate $K$ using the target error MSE\textsubscript{t} and gap. As shown in Table C1, when $K=3$, MSE\textsubscript{t} stabilizes and gap reaches its maximum. We also observe that greedy search outperforms individual variable selection.

\noindent
\textbf{Scaling Factor} ${{\lambda }_{\text{tim}}}$. In Design 1, $\lambda_{\text{tim}}$ is used to locate the target timing. We set its search range to [0,0.3], since positive deviations lead to larger operational damage, while larger amplification factors yield no further impact. This trend is shown in Fig. C1.

\begin{table}[t] 
\vspace{-0.6em}
    \centering 
    \caption{Analysis of poison variable K}
    \label{tab:K}
    \renewcommand{\arraystretch}{0.95}
    \setlength{\tabcolsep}{8pt} 
    \resizebox{\columnwidth}{!}{ 
    \begin{tabular}{ l| c c c c||c c}
        \toprule
        \textbf{} & K=1 & K=2 & \textbf{K=3} & K=4 & Top-K & Greedy \\
        \midrule
        MSE\textsubscript{t} ↓ & 0.067 & 0.050 & 0.043 & 0.042 & 0.049 & 0.043 \\
        gap ↑ & 0.224 & 0.251 & 0.278 & 0.278 & 0.255 & 0.278 \\
        \bottomrule
    \end{tabular}
    }
    \vspace{-0.3em}
\end{table}

\begin{figure}[t]
\centering
\includegraphics[width=\columnwidth]{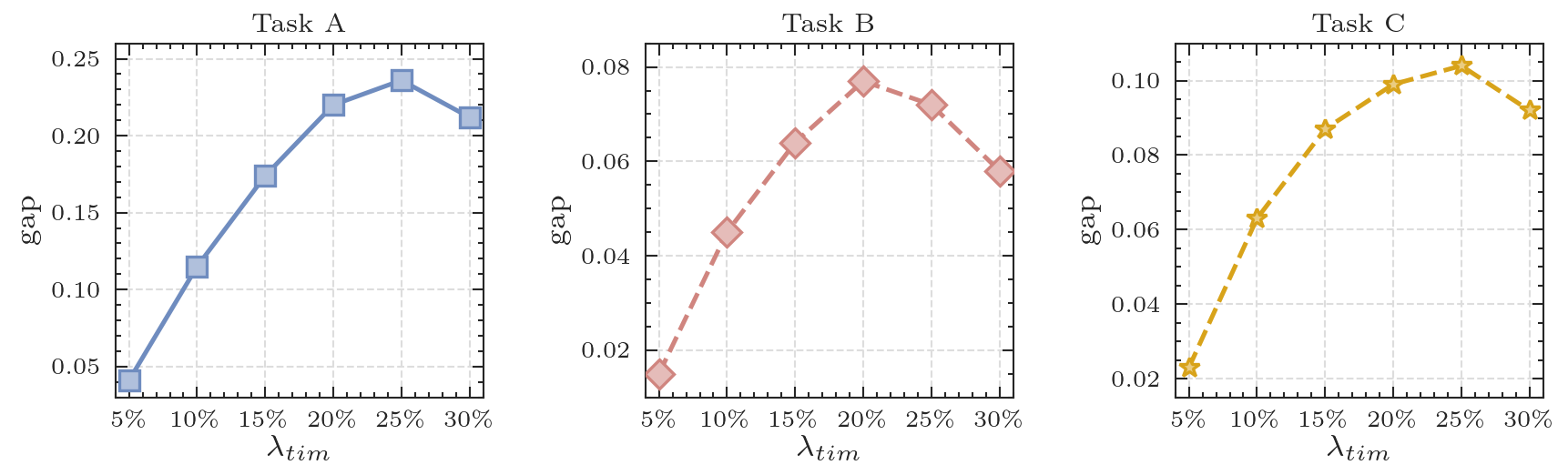}
\caption{Analysis of scaling factor ${{\lambda }_{\text{tim}}}$}
\label{fig:x}
\vspace{-0.5em}
\end{figure}

\vspace{0.2em}
\noindent
\textbf{Analysis of Multi-Trigger Strategy.} We argue that planting multiple triggers facilitates target fitting, especially for power system time series with strong daily periodicity. As shown in Table C2, the number of peaks varies across input lengths. In most cases, the number of selected triggers is consistent with the number of historical input days. The symbol `-' means that the corresponding peak does not satisfy the multi-trigger ratio threshold $\rho_{\text{mp}}$, and is therefore not selected.

\begin{table}[h] 
\vspace{-0.3em}
    \centering 
    \caption{Operation gap under different number of triggers}
    \label{tab:K}
    \renewcommand{\arraystretch}{0.95}
    \setlength{\tabcolsep}{7pt} 
    \resizebox{\columnwidth}{!}{ 
    \begin{tabular}{ l| c c c c c c c}
        \toprule
        \textbf{} & 1 & 2 & 3 & 4 & 5 & 6 & 7 \\
        \midrule
        96 & 0.231 & \textbf{0.243} & -- & -- & -- & -- & -- \\
        288 & 0.235 & 0.244 & \textbf{0.278} & -- & -- & -- & -- \\
        672 & 0.225 & 0.229 & 0.262 & 0.277 & 0.277 & 0.276 & \textbf{0.280} \\
        \bottomrule
    \end{tabular}
    }
    \vspace{-0.3em}
\end{table}

\vspace{0.2em}
\noindent
\textbf{Analysis of Poison Samples.} In Design 4, we poison two types of samples: those most similar to the target pattern and least similar from it. We study the effect of their ratio ${{\gamma }_{1}}:{{\gamma }_{2}}$, as shown in Fig. C2. GridTroj achieves the best performance when the two types are balanced at 1:1. When dissimilar samples are insufficient, MSE\textsubscript{t} increases sharply and gap decreases, indicating that dissimilar samples are important for learning a stronger trigger-target association.

\begin{figure}[h]
\centering
\includegraphics[width=\columnwidth]{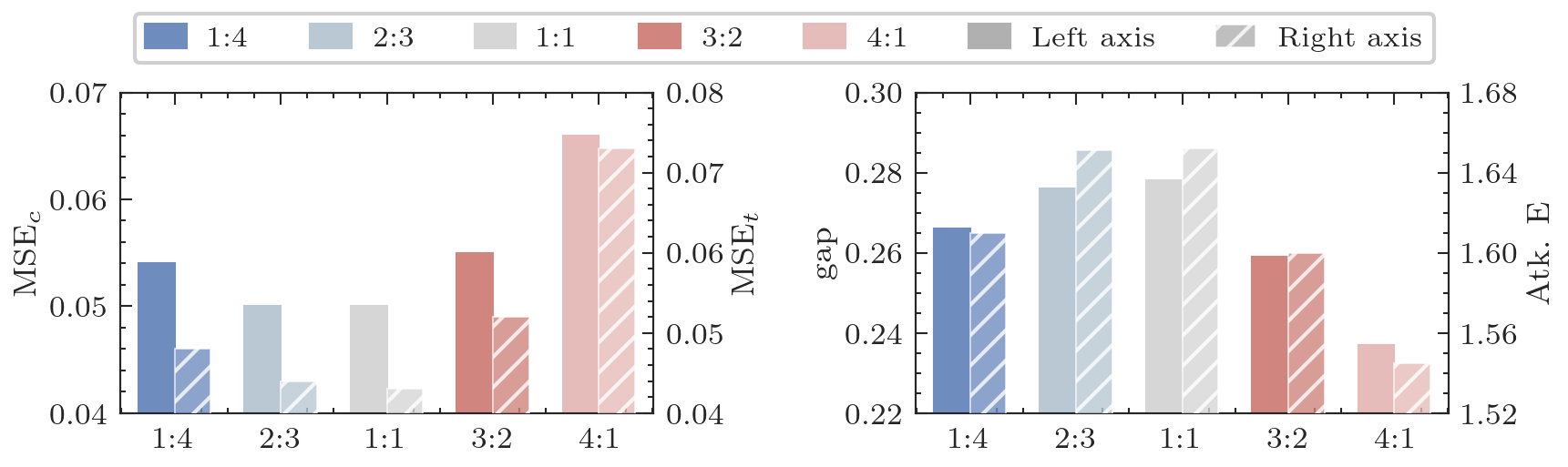}
\caption{Impact of the poison sample ratio ${{\gamma }_{1}}:{{\gamma }_{2}}$}
\label{fig:x}
\vspace{-0.3em}
\end{figure}

\noindent
\textbf{Implementation of Operation-Aware Loss.} In Module 4, the operation loss weight ${{\beta }_{2}}$ is set to 0.1, since a larger weight causes unstable training in some cases. We also find that introducing this operational loss strictly during the final 25\% epochs yields optimal convergence. Moreover, we compare the surrogate network approach against the numerical gradient obtained from the first-order Taylor expansion of the downstream optimization objective. The results show that when the samples are dense enough to cover the neighborhood of the target pattern, the surrogate network clearly outperforms the first-order gradient approximation. The trade-off is a higher but acceptable pre-training overhead.

\vfill

\end{document}